\newcommand{\Rs}{R_{\odot}}

\newcommand{\f}[2]{\frac{#1}{#2}}

\newcommand{\pder}[2]{ \f{\partial #1}{\partial #2} }
\newcommand{\grad}{ {\bf \nabla } }

\newcommand{\EQ}{\begin{equation}}
\newcommand{\EN}{\end{equation}}
\newcommand{\EQA}{\begin{eqnarray}}
\newcommand{\ENA}{\end{eqnarray}}
\newcommand{\bb}{\mbox{\boldmath $B$} {}}

\newcommand{\bm}{\boldmath}

\newcommand{\uu}{\mbox{\boldmath $u$} {}}

\newcommand{\nab}     {\nabla}

\documentclass[trackchanges,twocolumn]{aastex701}
\usepackage{bm}
\usepackage{amsmath}
\usepackage{booktabs}
\usepackage{hyperref}
\usepackage{mathtools, nccmath}
\usepackage[b]{esvect}

\usepackage{color}
\usepackage[normalem]{ulem}

\begin{document}

\title{Dynamics of Streamers and Pseudostreamers and Implications for the Solar Wind}

\author[orcid=0000-0002-3369-8471,sname='North America']{Sahel Dey}
\affiliation{School of Science, University of Newcastle, University Drive, Callaghan, NSW 2308, Australia}
\email[show]{sahel.dey@newcastle.edu.au}  
\author[orcid=0000-0002-1089-9270,gname=Bosque, sname='Sur America']{David I. Pontin} 
\affiliation{School of Science, University of Newcastle, University Drive, Callaghan, NSW 2308, Australia}
\email{david.pontin@newcastle.edu.au}

\author[orcid=0000-0003-0176-4312,gname=Savannah,sname=Africa]{Spiro K. Antiochos}
\affiliation{Department of Climate and Space Sciences and Engineering, University of Michigan, Ann Arbor, MI 48109, USA}
\email{spiro.antiochos@gmail.com}
\begin{abstract}

%
%
The origin of the Sun's slow wind and its inherent variability remain unknown, but there is increasing evidence that interactions between closed and open magnetic flux in the corona play a major role.
%
%
This paper studies the dynamic evolution of streamers and pseudostreamers with a particular focus on the release of plasma from the closed to the open field region.
%
%
We employ a global 2.5D magnetohydrodynamic model that extends from the solar chromosphere to 30 solar radii, and that extends previous interchange magnetic reconnection modelling by including self-consistent thermodynamics.
%
%
We find that at both the helmet streamer and pseudostreamer there is a continual dynamic interaction between closed and open flux even in the absence of explicit driving. At the helmet streamer, the dynamics take the form of a ``breathing'' cycle in which the closed flux contracts and expands, and plasmoids are released along the heliospheric current sheet.  The pseudostreamer exhibits a back-and-forth motion, driving interchange reconnection alternately on its opposite flanks. The resulting release of hot, dense plasma leads to density fluctuations in the open field that are significantly larger above the helmet streamer due to the persistence of the plasmoids there. 
%
%
Our model demonstrates that plasma is continually being released into the heliosphere from both streamers and pseudostreamers. We discuss the implications of our simulation results for observations of the corona and inner heliosphere.
\end{abstract}



\section{Introduction}
The solar wind is a stream of plasma that continuously outflows from the Sun and fills the heliosphere. The existence of the solar wind was first proposed by \cite{parker1958} with a steady-state, isothermal model. Later, the Mariner 2 spacecraft confirmed the signature of the solar wind using in-situ measurements \citep{neugebauer1962}. The observations made from satellites such as Ulysses further reveal that the solar wind has two different wind streams classified as ``fast'' and ``slow'' solar wind \citep[e.g.,][]{mccomas2008}. The fast solar wind originates mostly from the 
the central regions of
coronal holes where magnetic fields are open {(i.e., the magnetic field lines from these regions extend far out into the heliosphere)}. It has a typical radial speed greater than 500 km\,s$^{-1}$ when measured around 1 AU. 
In recent times, the Solar Orbiter and Parker Solar Probe (PSP) missions have provided a plethora of information on wind composition and variability using limb observations and in-situ measurements, respectively. For a review relating to the solar wind and its connection to the Sun's corona, see \cite{cranmer2019}.

In this paper, we undertake modeling aimed at increasing our understanding of the slow wind. A detailed review of observations and theories relating to the slow wind is given by \cite{abbo2016}.
The precise origin of the streams of slow wind remains to be determined, but there is growing evidence that they typically originate from the boundaries of coronal holes \citep[often around active regions; e.g.,][]{owens2013,brooks2015,Arge2024}, where open magnetic flux abuts closed magnetic flux. Thus, one proposed mechanism for the formation of the slow wind involves the interaction of closed and open magnetic flux through interchange magnetic reconnection \citep{Suess96,fisk1998,antiochos2011}.
As shown by \cite{antiochos2011}, open magnetic flux passing close to closed flux in the low corona can fill a substantial portion of the heliosphere as the so-called ``S-Web''. This S-Web is associated with different magnetic structures in the corona that are consistent with the helmet streamers and pseudostreamers found in observations \citep[e.g.,][]{scott2018}.  Recently, \cite{wilkins2025} demonstrated that the percentage of heliospheric flux that is rooted in the photosphere within 25~Mm of the closed field varies between around 90\% at solar maximum and around 40\% at solar minimum, consistent with the solar-cycle variation of fast and slow wind fractions \citep{tokumaru2021}.

Computational models of the solar wind are based on a range of different physical models, depending on the target physics. The goal is to provide sufficient thermal pressure in the low corona to drive a solar wind outflow, with some desired characteristics that match observations, such as the radial outflow speed.
For all models on these global scales, it is impossible to resolve the (yet-to-be-determined) processes that heat the open and closed magnetic field regions -- such as nanoflare reconnection or wave damping mechanisms. Instead, some source term is usually added to the energy equation.
One common approach (previously implemented, for example, in the MPI-AMRVAC code that we use in this paper) is to control the wind speed by setting a target temperature and adding (or subtracting) energy from the plasma to maintain this temperature. This approach has been used, for example, to study propagation of coronal mass ejections (CMEs), where the goal is simply to have a wind profile (as a function of radius and latitude) that resembles observations in order to transport these CMEs outwards \citep[e.g.,][]{jacobs2005}. Other computational approaches seek to reproduce the observed properties of the global corona and solar wind in more detail. For example, \cite{lionello2009s} present global MHD simulations wherein they seek to reproduce the emission from the Sun across a range of EUV and X-ray wavelengths. They compared a number of different heating terms including a term decaying exponentially with radius to provide a background heating that -- when coupled with an additional pressure term in the momentum equation -- can provide a fast wind outflow. They also included two magnetic field-dependent terms to account for Quiet Sun and Active Region heating. The latter provides extra heating in strong field regions to match the observed intensity enhancement in active regions \citep[see also][]{downs2010}. As well as these empirical heating models, some studies use physically motivated heating models, such as those based on Alfv\'en wave turbulence \citep[e.g.][]{sokolov2013,vanderholst2014,reville2020}. One advantage of such models is that they can reproduce the different heating rates in closed and open field regions self-consistently. Finally, while all of the above-mentioned models have a lower (radial) boundary somewhere above the solar surface, \cite{iijima2023} presented a model for an angular wedge that included the sub-surface convection zone out to the solar wind.

Our goal here is to explore the dynamics of helmet streamers and pseudostreamers, particularly in the context of the release of plasma from the closed field to the open field by interchange reconnection. We build on previous work studying the plasma and magnetic field dynamics in the vicinity of the open-closed boundary, which has addressed the dynamics within narrow corridors of open flux \citep{higginson2017b}, the motion of isolated parasitic polarity regions \citep{edmondson2010,pontin2013}, the formation of a pseudostreamer in the open field by migration from the closed field \citep{scott2021}, and the impact of supergranular driving at helmet streamer and pseudostreamer boundaries \citep{aslanyan2021,aslanyan2022}. Most of these studies were undertaken using the ARMS code, and all were restricted to an isothermal atmosphere/wind. In particular, \cite{aslanyan2021,aslanyan2022} showed that photospheric supergranulation flows drive the formation of interchange-reconnected field lines in narrow filaments. The implication is that the outflow of closed-field plasma should form narrow channels even away from the photosphere, consistent with the imprint of the supergranulation detectable out in the solar wind \citep{borovsky2016}.
In this paper our goal is to present a model that extends these studies by including a self-consistent thermodynamic evolution of the plasma. This is critical for comparing plasma signatures in the model with those observed by, for example, Solar Orbiter and Parker Solar Probe.

The structure of the paper is as follows. In Section \ref{sec:method} we describe the physics included in our model and the computational method.  In Sections \ref{sec:steadystate}--\ref{sec:dynamics} we present the results of our simulations. 
In Section \ref{sec:in-situ} we compute synthetic observable and discuss the relevance for observations, and 
in Section \ref{sec:discuss} we finish with a discussion.

\section{Method}\label{sec:method}
\subsection{Governing equations and algorithms}
We solve the MHD equations
using the MPI-AMRVAC code~\citep{porth2014,xia2018,keppens2023}. MPI-AMRVAC is a highly modular finite volume Magnetohydrodynamics framework, suited for solving hyperbolic and parabolic partial differential equations associated with various astrophysical phenomena and beyond~\citep{keppens2020}. In our numerical setup, we define the domain with a 2.5D spherical-polar mesh that allocates all variables on a constant azimuthal plane but includes three components of all vector fields. The radial extent of the domain ranges from the {low} solar chromosphere, located at $1 R_{\odot}$, to the outer solar corona at $30 R_{\odot}$. The locations of the grid cell centers are specified on a radially stretched mesh, which increases in radial separation following a geometric progression~\citep{xia2018}. Further, we utilize seven levels of adaptive mesh refinement on the base stretched grid to include extremely small and large scales at the same time. Our refinement strategy is similar to a Lohner-type prescription~\citep{Lohner1987}, which is based on computing second-order derivatives weighted by specified variables. Specifically, we set 30\% weight on each radial and poloidal component of the magnetic field and 20 \% weight  each on the density and internal energy. When mesh refinement occurs, it acts in both the radial and poloidal directions to preserve the primary aspect ratio of the cell. In our domain, the smallest and largest radial grid spaces are approximately $250$ km and $0.78~\Rs$, respectively, under the adaptive mesh refinement setting. The poloidal range linearly covers from the north to the south polar region, with a maximum poloidal grid spacing of $2.81^{\circ}$ and a minimum of $0.02^{\circ}$.

Our simulations solve the following mass continuity, momentum conservation, magnetic induction equation and internal energy equation of the plasma medium: 
\begin{equation}
    \pder{\rho}{t}+\grad \cdot (\rho \uu) =0,
\end{equation}
\begin{equation}
    \pder{\left(\rho \uu\right)}{t}+\grad \cdot \left(\left(p_{g}+p_{m}\right)\mathbf{I}+\rho\uu\uu-\mathbf{BB}\right) =\rho \mathbf{g},
\end{equation}
\begin{equation}
    \pder{\bb}{t}+\grad\cdot(\uu\bb-\bb\uu)=0,
\end{equation}
\begin{equation}\label{eq:heat}
    \pder{e}{t}+\grad\cdot(e\uu)+p_g\grad\cdot\uu=H-\grad\cdot\mathbf{Q}-n^{2}\Lambda(T),
\end{equation}
where $\rho$ is the mass density, ${\bf u}$ is the plasma velocity, $p_g$ is the gas pressure, $p_m$ is the magnetic pressure, ${\bf B}$ is the magnetic field and $e$ is the internal energy density. For the equation of state, we use the Ideal gas law to define $p_g$. $\mathbf{g}$ represents the solar gravitational acceleration, which has a maximum value of $274$m\,s$^{-2}$ at the lower boundary and follows a $1/r^{2}$ profile with radius. $H$ and ${\bf Q}$ refer to the coronal heating source and anisotropic conductive heat flux, respectively. Both terms are described comprehensively in Section~\ref{sub:Energetics}. The effect of the optically thin radiative cooling is included by the $n^{2}\lambda(T)$ term, where $n$ represents electron number density and $\lambda(T)$ is the radiative cooling function. We selected a cooling function based on solar coronal abundances~\citep{dere2009}.

To solve the coupled PDEs in the conservative approach, we choose an HLL-type approximate Riemann solver to discretize the numerical domain and a third-order-accurate three-stage Runge-Kutta method for the time integration scheme~\citep{Gottlieb1998}. 
The magnetic field is written as the sum of the initial magnetic field plus a (general, non-linear) perturbation, with only the perturbation field being evolved.
This approach ensures better numerical stability -- particularly in the low plasma-$\beta$ regime~\citep{Tanaka1994} -- and the associated modifications in governing equations are described thoroughly in \cite{xia2018}. The spatial reconstruction from cell center to cell interface for the flux computation is performed using the Vanleer limiter~\citep{vanleer1974} with second-order accuracy. To maintain the divergence-free constraint for the magnetic field, we opt for both the diffusion~\citep{keppens2003} and additional source term approach~\citep{powell1999} in our model.
\subsection{Initial conditions and boundary conditions}
Our aim is to model a system in which the plasma in the magnetically closed regions is hotter and denser than in the open field regions, as observed on the Sun. Because the identities of closed and open field lines will change over time, this should occur self-consistently within the model. This necessitates the presence of a chromospheric layer at the base of our domain, in order to supply the necessary evaporative upflows. Thus, our
initial temperature profile starts from 24,500 K at the chromosphere, then includes a transition region and a 1.25 MK hot corona. At $t=0$ we impose the temperature profile
\begin{equation}
T(r)=T_b\left(b+\frac{1}{2}\left(1+\tanh{\left(\frac{r-h}{a}\right)}\right)\right),
\end{equation}
where $T_b=1.23$~MK, $b=2\times10^{-2}$, $h=1.007~R_{\odot}$ and $a=10^{-3}~R_{\odot}.$
Using this temperature variation, we solve the 1D-hydrostatic balance equation under solar gravity as discussed above and determine the initial density stratification in terms of radial distance. The chromospheric density is set to $3.32\times10^{-13}{\rm \,g\,cm}^{-3}$ at the bottom of the domain. The initial velocity field is fixed at zero. 

In this model, our goal is to study helmet streamer and pseudostreamer structures in the solar corona. Thus, to initialise the magnetic field ($\mathbf{B}_{\rm ini}$) in the simulations, we take a superposition of a global dipolar field {($\mathbf{B}_{\rm dip}$)} 
with an additional field ($\mathbf{B_{\rm add}}=\grad\times \mathbf{A_{\rm add}}$) that contributes only in the following co-latitude($\theta$) range; 
\begin{equation}
\mathbf{B}_{\rm ini}=\mathbf{B}_{\rm dip}+\mathbf{B_{\rm add}} ;\quad 1.02~(58.48^{\circ})\le\theta\le 2.52~(144.43^{\circ})
\end{equation}
Note that $\mathbf{B_{\rm add}}$ has been constructed so that it vanishes at the end points of the co-latitude range above. Outside this $\theta$ range, we take only the global dipole ($\mathbf{B}_{\rm dip}$) component to contribute to the initial field ($\mathbf{B}_{\rm ini}$).  
\begin{equation}
\mathbf{B}_{\rm ini}=\mathbf{B}_{\rm dip}
\end{equation}
\begin{equation}
\mathbf{B_{\rm dip}}=\frac{2B_{0}\cos{\theta}}{r^3}\hat r+\frac{B_{0}\sin{\theta}}{r^3}\hat \theta
\end{equation}
where $B_0=-3~G\cdot R^{3}_{\odot}$. The additional magnetic field is defined in terms of the vector potential ($\mathbf{A_{\rm add}}$)~\citep{talpeanu2022}, via       
\begin{equation}
\mathbf{A_{\rm add}}=\frac{A_{0}}{r^{4}\sin{\theta}}\cos^{2}{\left(\frac{\pi}{2\delta a}\left(\frac{\pi}{2}-\theta+\theta_{0}\right)\right)}\hat\phi
\end{equation}
where $A_{0}=2.60~G\cdot R^{5}_{\odot}$, $\theta_{0}=0.2~(11.45^{\circ})$ and \\
$\delta a= 0.75~(42.97^{\circ})$.
At the lower radial boundary, the density and temperature are fixed to their initial values for the duration of the simulation. The bottom boundary is closed for any flows. Both the mass density and radial momentum density are proportional to $1/r^2$ at the outer radial boundary, where the temperature follows a $1/r$ profile. The magnetic field is set as line-tied at the lower radial boundary. {At the top boundary, radial, poloidal and azimuthal components of the total magnetic field follow $1/r^{2}$, constant and $1/r$ profiles, respectively.}
\subsection{Energetics}
\label{sub:Energetics}
The thermal heat transport in our model is determined in general by
the magnetic field-aligned thermal conduction coefficient ($k_{\parallel}$). For the collisional plasma regime, it is defined in Equ.~\ref{eq:heatflux}, where $ k_{\parallel}=10^{-6} T^{2.5}~{\rm erg \cdot cm^{-1}\cdot s^{-1} {\cdot K^{-1}}}$. 
To mimic the effect of reduced collisionality with increasing radius, the conduction coefficients are damped following a $1/r^{2}$ dependency ~\citep{shoda2019,matsumoto2021}. This is an effectively similar approach to considering the free stream conduction flux for the non-collisonal plasma medium, as the heat is mostly trapped within plasma elements in the absence of strong conduction and the conduction speed is determined by the motion of the plasma elements. Specifically, we let the heat flux(${\bf Q}$) in Equation (\ref{eq:heat}) take the form
\begin{equation}
     \mathbf{Q}=-k_{\parallel}\hat b\hat b\cdot\nab T
     \label{eq:heatflux}
\end{equation}
Where $\hat b$ is the unit vector along the magnetic field.

To sustain the million Kelvin order temperature of the solar corona, there are several possible mechanisms to consider in numerical models, e.g., Alfv\'enic turbulence, random heating, and damping of magnetohydrodynamic waves. In our setup, we choose a simple volumetric heating function ($H$) that has been used in a number of prior studies~\citep{Fan2017,Gannouni2023,Maity2024,Singh25}: 
\begin{equation}\label{eq:heatingfn}
    H=\left(\frac{R_{\odot}}{r}\right)^2 \left(H_{0}\exp{\left(-\frac{r-r_0}{\lambda_0}\right)}+H_{1}\exp{\left(-\frac{r-r_0}{\lambda_1}\right)}\right).
\end{equation}
Here $H_{0}= {2.86\times10^{-6}~\rm erg\cdot cm^{-3}\cdot s^{-1}}$, $\lambda_{0}=0.7$ $R_{\odot}$ Mm, $H_{1}= {5.72\times10^{-5}~\rm erg\cdot cm^{-3}\cdot s^{-1}}$, $\lambda_{1}=40$ Mm and $r_0=1 R_{\odot}$.
The first term of the heating function contributes to maintaining the coronal temperature on a global scale~\citep{Gannouni2023} by introducing total power of $8.44\times10^{27}$ $\rm erg\cdot s^{-1}$. The second source term generates the relatively hot closed magnetic fields compared to the open field regions~\citep{lionello2009s}, which are frequently detected in coronal observations. The total injected power for this comparatively short length scale heating source is  $1.37\times10^{28}$ $\rm erg\cdot s^{-1}$.

It is well established that maintaining realistic energy fluxes across the solar transition region is hugely challenging computationally, due to the abrupt change in temperature from the chromosphere to the corona. Under-resolving the transition region leads to an underestimation of the conductive energy flux, with the result that the energy balance does not match the true coronal conditions. To mitigate these issues, we employ the Transition Region Adaptive Conduction method \citep[TRAC; ][]{jhonston2019,jhonston2020} to artificially broaden the transition region while maintaining the appropriate energy balances. Specifically, we use the TRAC implementation from ~\cite{jhonston2020} in MPI-AMRVAC.


\section{Plasma Properties of the Quasi-equilibrium State}\label{sec:steadystate}
\begin{figure}
    \centering
    \includegraphics[width=0.9\linewidth]{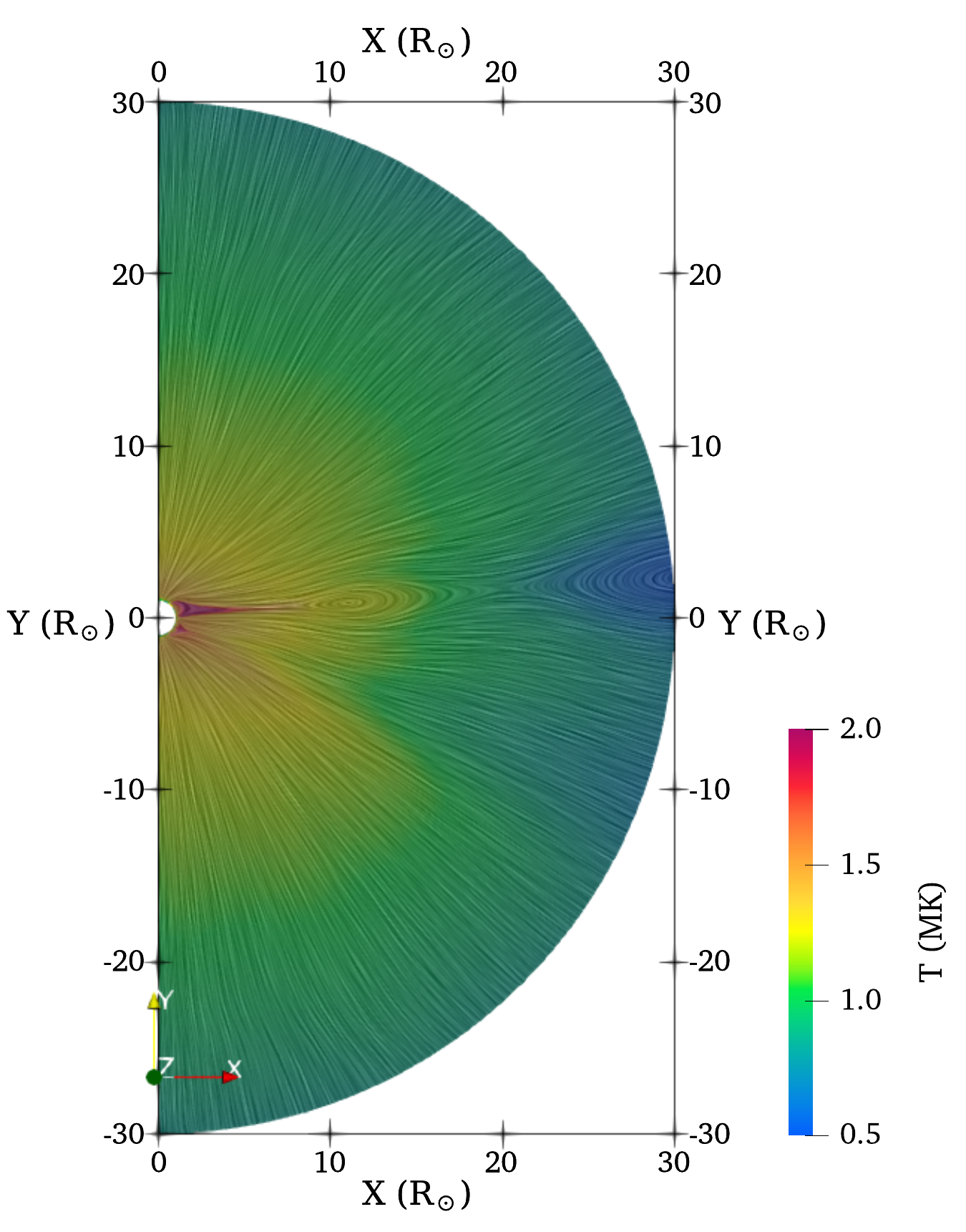}
    \caption{A global-scale plasma temperature map from the quasi-equilibrium state, at $t= 80$ hours of solar time. The associated magnetic field lines are visualized by the Line Integral Convolution (LIC) method.}
    \label{fig:globalmap}
\end{figure}
\begin{figure*}
    \centering
    \includegraphics[width=\linewidth]{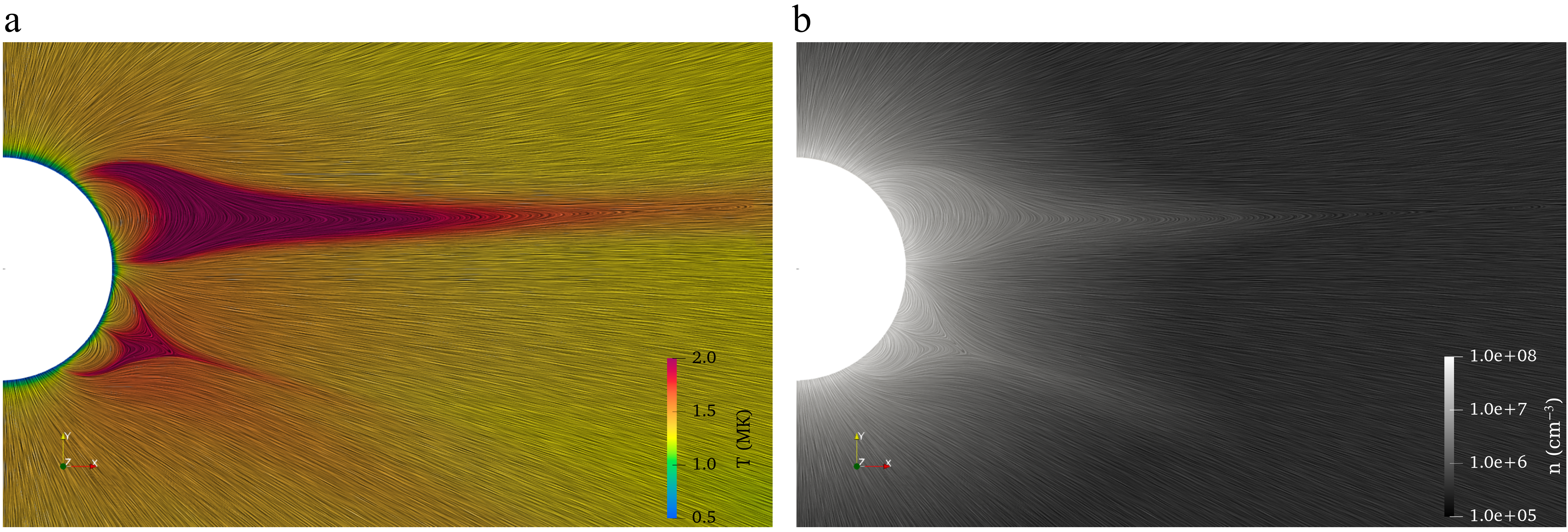}
    \caption{Temperature and density structures of the simulated Helmet streamer and Pseudostreamer at the quasi-equilibrium state at $t=80$ hours. Panel (a) is a close-up view of Figure \ref{fig:globalmap} showing the temperature, while panel (b) shows the number density profile.}
    \label{fig:streamers}
\end{figure*}
\begin{figure}
    \centering
    \includegraphics[width=0.9\linewidth]{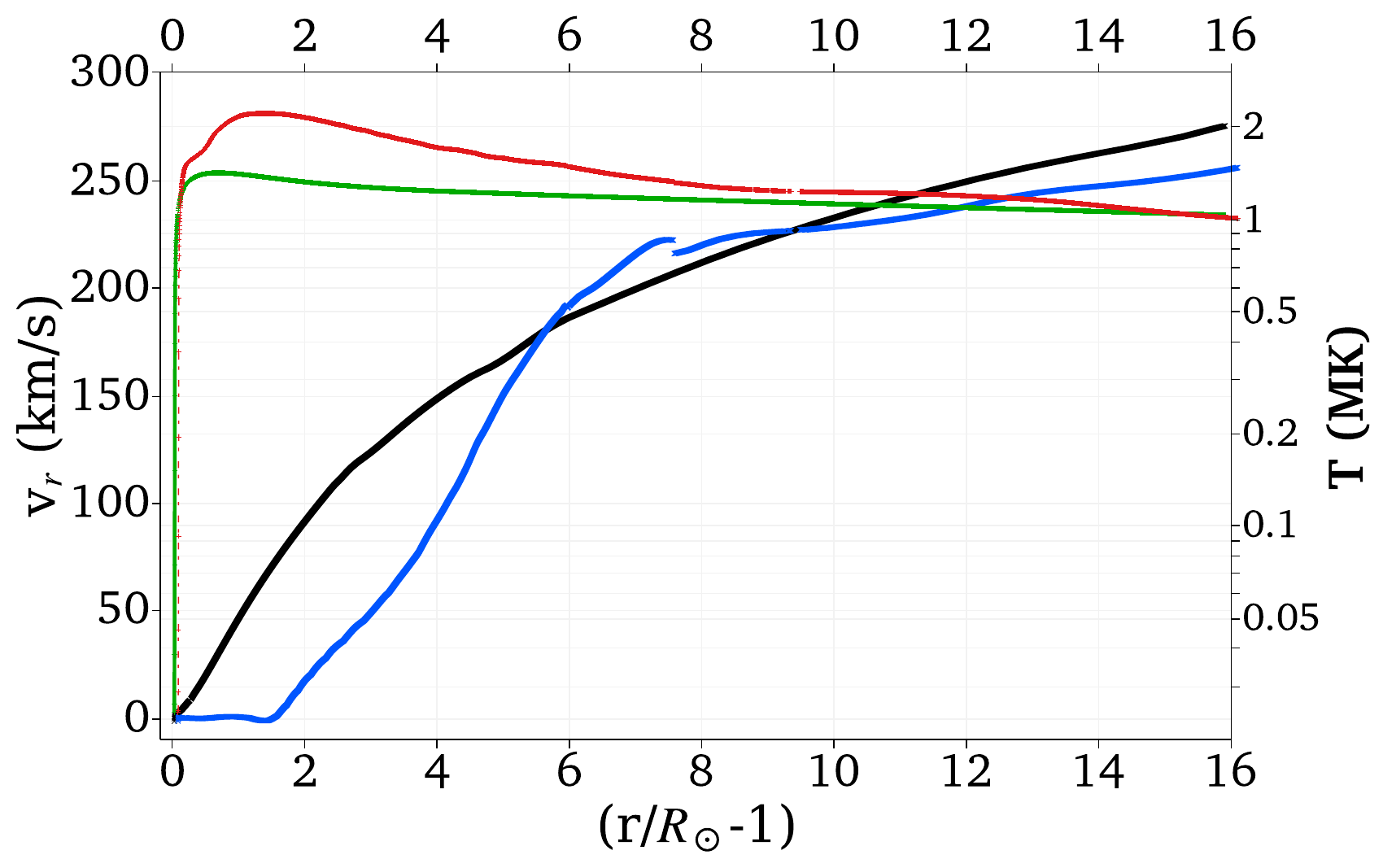}
    \caption{Variation of the radial flow speed and temperature of the solar wind with height (at t=80 hours). Black and blue curves represent radial outflow speed (left axis) measured near the north pole and along the helmet streamer stalk, respectively. Temperature profile (right axis) near the polar region and along the same streamer stalk are shown with green and red lines, respectively.}
    \label{fig:radial_profiles}
\end{figure}

Using the simulation setup and physics described above, we evolve the system from the initial state until it reaches a quasi-steady state at 80 hours of solar time. Until $t=20$ hours, we restrict the maximum refinement to five levels while multiple large-scale shock waves pass through in response to the strong initial force-imbalanced state. As shown in Figures \ref{fig:globalmap} and \ref{fig:streamers}(a), the quasi-equilibrium state exhibits helmet streamer and pseudostreamer structures at Northern and Southern latitudes, respectively. 
The temperature distribution of the whole domain at t=80~hours is shown in Figure \ref{fig:globalmap}, which has many localized structures in contrast to the initial uniform 1.25 MK temperature (in the coronal part of the domain). It is clearly noticeable that the plasma near the magnetic poles is cooler than the closed magnetic field regions of both streamers.  At $r=2$\,R$_\odot$, the temperature at the north pole is 1.39 MK, where the closed field regions of the helmet streamer and pseudostreamer are at 2.16 MK and 1.65 MK, respectively.  In Figure.~\ref{fig:radial_profiles}, the temperature profiles further support this contrast between the polar region and the Helmet streamer out to $r=10~R_{\odot}$. 

It is important to mention that the rate of energy-density injection as the coronal heating process (discussed in Section~\ref{sub:Energetics}) is the same for all latitudes at a given radius. Therefore, this temperature contrast between the open and closed magnetic field regions is due to the well-known difference in the primary cooling process between closed and open flux regions. In the closed field regions the heating into the corona must be balanced by radiative losses. Thermal conduction serves only to redistribute the energy so that a substantial fraction of the radiative losses are from the transition region \citep[e.g.][]{jhonston2019}. As a result, the temperature and especially the density in the closed corona must be high enough to supply the needed losses. In the open field, however, the heating can be balanced by the large enthalpy flux of the solar wind. As a result, the radiative losses are substantially lower, implying lower temperatures and densities. This strong density contrast between the open and closed can be seen in Figure.~\ref{fig:streamers}(b): for example, at $3R_\odot$, the typical helmet streamer density is 7.31 times larger than in the polar region. This same contrast approximately holds even at higher distances (e.g., $r=10~R_{\odot}$) when measured above the streamer stalk. Note that these results are completely in line with the typical observations.

In terms of the plasma kinematics, solar wind streams are developed from the initial zero-velocity state and sustained self-consistently. In our model, coronal gas pressure drives the outflow as the wind. For a Sun-like G-type star, it is believed to be the primary source of wind generation~\citep{Lamers1999}. In observations, the solar wind speed varies significantly with solar latitude \citep[e.g.,][]{mccomas2008}. We observe a similar qualitative behavior of the wind speed profile in our model. At $15$\,R$_\odot$, our wind reaches 261 km/s radial speed near the polar regions (see Figure \ref{fig:radial_profiles}). At a similar height, the wind above the Helmet streamer stalk reaches approximately 245 km/s. In a distance range between $7R_{\odot}$ and $10~R_{\odot}$, plasma from the streamer stalk outflows faster than the polar counterpart. This is because of a propagating local eruption from the helmet streamer, which we will discuss in subsection~\ref {subsec:dyn_helmet}. The solar wind is further accelerated and reaches a speed of 320 km/s at $30$\,R$_\odot$. Inside the closed field structures of both streamers, plasma motion is dictated by the magnetic field geometry; hence, there are no signatures of the wind. 
However, continuously fluctuating field-aligned flows are present, with speeds up to 8 km/s in multiple magnetic loops of both streamers. We argue that these field-aligned flows, together with the plasma thermodynamics, are responsible for the dynamic evolution, particularly of the pseudosteamer, discussed in the following section.
\newline

After reaching the quasi-steady state, we introduce a shearing velocity profile near one of the legs of each of the streamers on the lower boundary.
The aim of the shearing velocity perturbation in the azimuthal direction ($v_{\phi}$) is to simulate the effects on helmet and pseudostreamer and dynamics of the supergranulation motions, which add free energy into the coronal magnetic field. The driving shearing profile is constant in time and applied for approximately four supergranulation life cycles (96 hours). The explicit form of the perturbation for the helmet streamer ($v^{h}_{\phi}$) is defined in Equ.~(\ref{eq:shear1})
\begin{equation}
\label{eq:shear1}
\begin{split}
   v^{h}_{\phi}=v_0 \sin{\left(\frac{\pi\left(\theta-\theta_1+\delta\theta_1\right)}{2\delta\theta_1}\right)};\quad \theta_1-\delta\theta_1 \leq \theta\leq \theta_1+\delta\theta_1,
\end{split}
\end{equation}
where $v_0$, $\theta_1$ and $\delta\theta_1$ represent the amplitude of the shear, location of the helmet streamer foot point in the quasi-steady state at t=80~hours, and width of the shear profile, respectively. We set $v_0=0.5$ km/s, $\theta_1=44.1^\circ,\, \delta\theta_1=1.71^\circ$. Similarly, for the pseudostreamer, the shear velocity ($v^{p}_{\phi}$) is described in Equ.~(\ref{eq:shear2})
\begin{equation}
\label{eq:shear2}
\begin{split}
   v^{p}_{\phi}=v_0 \sin{\left(\frac{\pi\left(\theta-\theta_2+\delta\theta_2\right)}{2\delta\theta_2}\right)};\quad \theta_2-\delta\theta_2 \leq \theta\leq \theta_2+\delta\theta_2,
\end{split}
\end{equation}
where, $\theta_2=146.1^\circ$, $\delta\theta_2=1.71^\circ$. 
The effect of this velocity perturbation is to induce a non-zero magnetic field in the azimuthal direction at both streamer structures. This process will increase the magnetic free energy in the system, 
which can propagate freely outwards in the open field, but will be trapped and build up in the closed field region, building up a magnetic shear across the open-closed boundary.

\section{Dynamics of the Helmet Streamer and Pseudostreamer}\label{sec:dynamics}
\begin{figure*}
    \centering
    \includegraphics[width=\linewidth]{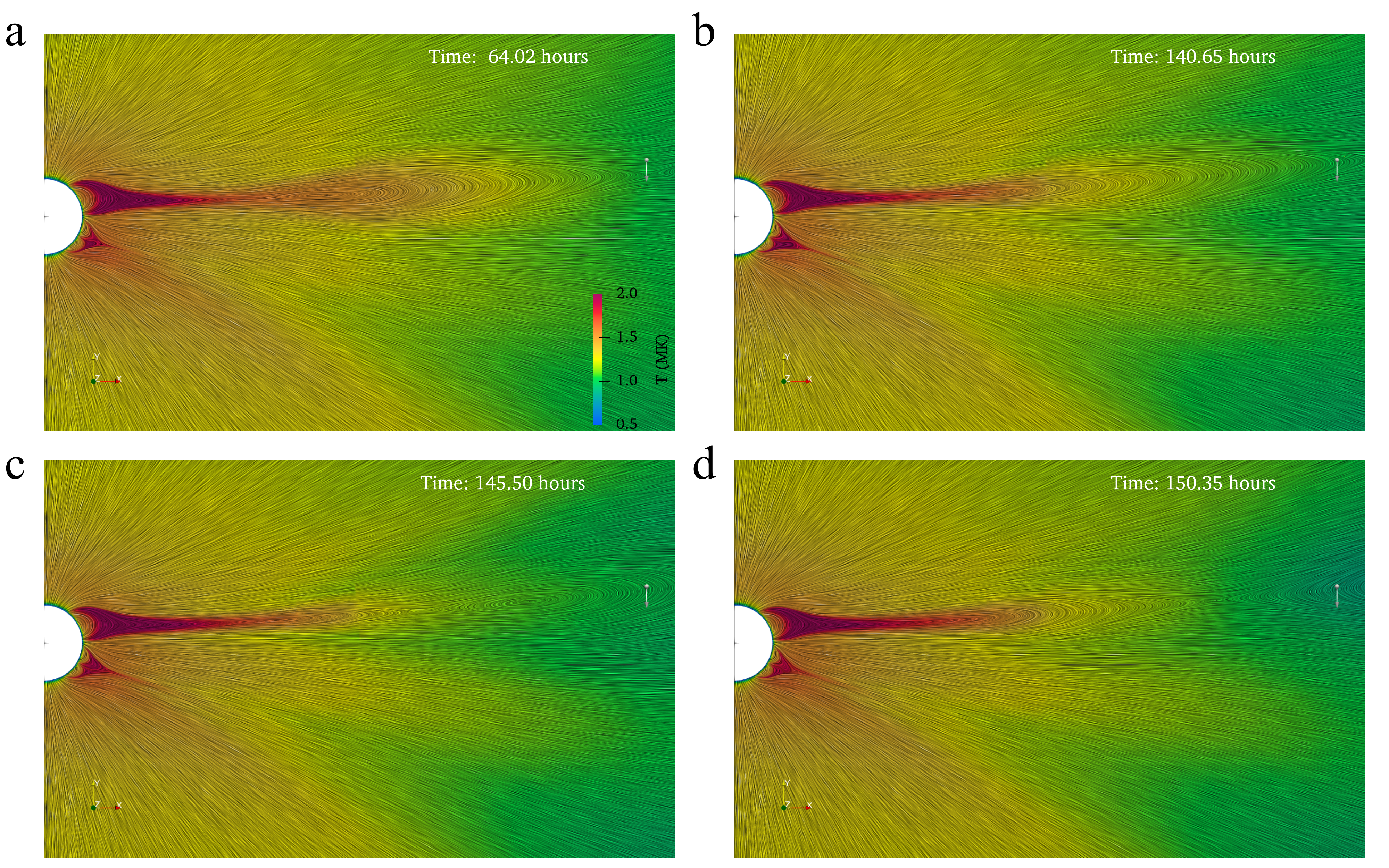}
    \caption{Quasi-periodic dynamics of the Helmet streamer: The temperature map from the quasi-equilibrium state of the Helmet streamer is shown in panel (a). An arrow is located at $r=16$~R$_\odot$ to show the location for synthetic in-situ measurements, discussed in Section~\ref{sec:in-situ}. Panels (b)--(d) represent different phases of streamer dynamics after introducing the shear in the azimuthal $(\phi)$ direction.}
    \label{fig:dhelmet}
\end{figure*}
\begin{figure*}
    \centering
    \includegraphics[width=0.9\linewidth]{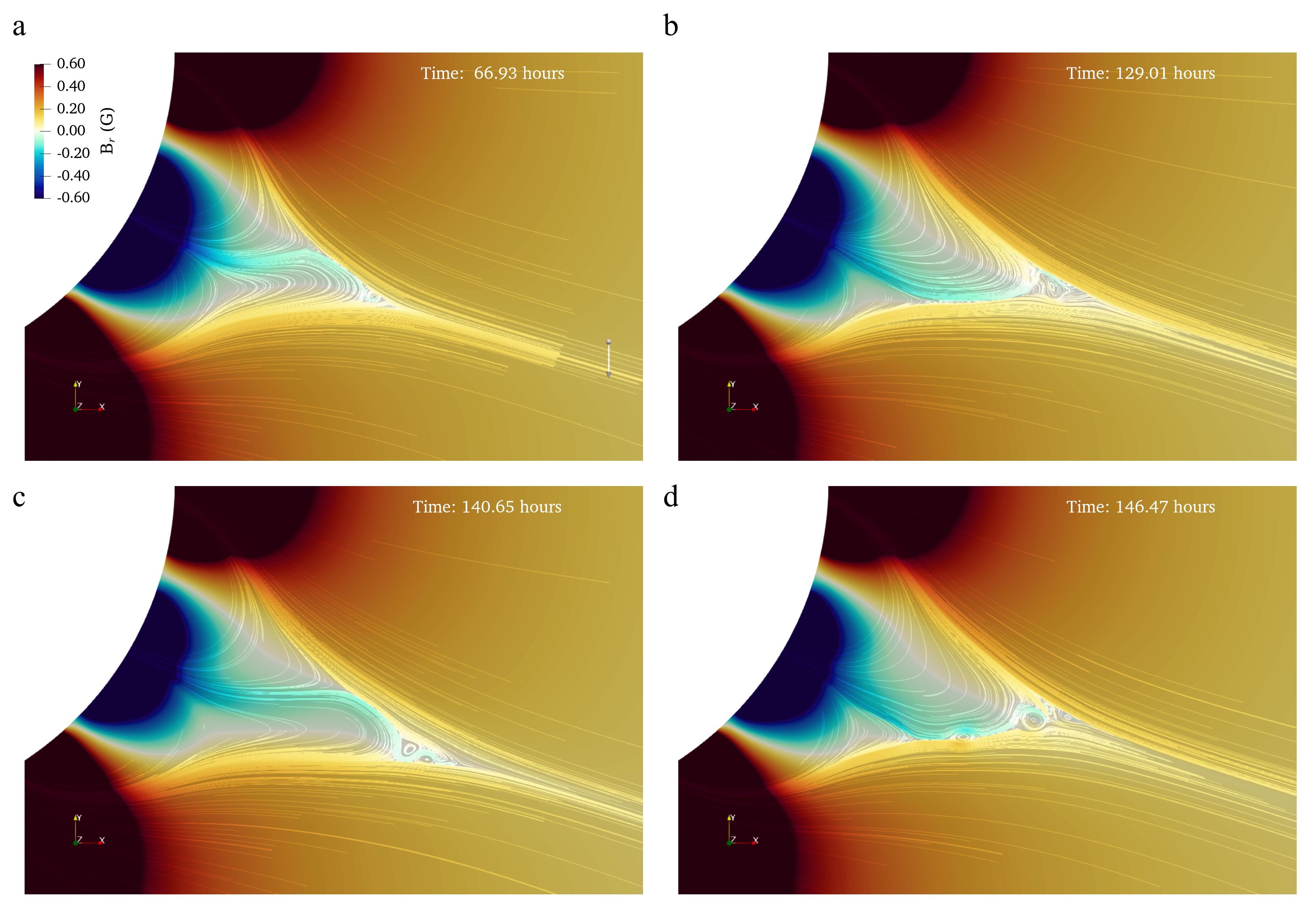}
    \caption{Dynamics in the vicinity of the pseudostreamer, revealing the presence of  Interchange Magnetic Reconnection (IMR). The radial component of the magnetic field is represented by the colour map and the magnetic field lines are traced from random seed points. Similar to Fig~\ref{fig:dhelmet}, the quasi-equilibrium state is shown in panel (a) and panels (b)--(d) refer to different times during the time period in which the shear is applied. An arrow is located at $r=2.5$~R$_\odot$ for a reference to compute synthetic in-situ observables (see Section \ref{sec:in-situ}).}
    \label{fig:dpseudo}
\end{figure*}
\subsection{Helmet streamer dynamics}\label{subsec:dyn_helmet}
{The intrinsic non-steady nature of the helmet streamer is clearly evident during both the quasi-equilibrium and driving phases. A snapshot of the streamer structure from the quasi-steady state is presented in Figure.~\ref{fig:dhelmet}a, while (b)--(d) panels show different times during the period when the shear driving is applied.} The apex of the closed field lines of the helmet streamer in the quasi-steady state lies around $5~R_\odot$. However, this height is constantly fluctuating, typically between $5~R_\odot-6~R_\odot$, while sometimes reaching even larger radii.
 It is in part modulated by periodic magnetic reconnection events in the overlying Helmet streamer stalk (i.e., in the heliospheric current sheet or HCS). This process leads to the formation of outgoing plasmoid structures as well as sunward reconnection flows or at least decrements in the local outflow speed.  
After each of these pinching-off reconnection events, the closed magnetic loops of the streamer shrink down to a lower height. With time, gas pressure again builds up in the closed field region due to the heating process and the streamer expands.
Subsequently, the solar wind acts to stretch out and eventually open up these closed field lines.
The net effect is a ``breathing'' cycle of the Helmet streamer in which the closed flux contracts and expands, and plasmoids are released along the HCS. The period of this ``pinching off'' of {plasmoids in our simulations is between $11.64-14.55$ hours},
and leads to a similar periodicity in density fluctuations in the vicinity of the HCS. This type of continuous plasmoid formation has been seen in other 2.5D \citep{allred2015} and 3D \citep{higginson2018} simulations of the quasi-steady HCS, even without the photospheric driving. Furthermore, recent PSP observations very near the Sun, $\sim 10 R_{\odot}$, suggest that the HCS is constantly reconnecting and generating plasmoids \citep{eastwood2026}.

To understand the effect of numerical resolution on the result, we have run an additional simulation in which we restrict the grid refinement to only 5 levels. We find a similar qualitative behaviour in terms of the plasmoid eruptions along the HCS. In the simulation with a coarser grid, the periodicity of the eruption increases compared to the standard simulation (with seven levels of refinement): likely, the reconnection onset is earlier due to the lower effective Lundquist number.
In terms of the plasmoid dynamics following their formation, these structures propagate radially outward with the background solar wind while at the same time expanding. Sophisticated observations from the WISPR instrument of the PSP clearly demonstrate the expansion of bright ellipsoidal streamer blob structures (plasmoid in 3D) as they accelerate outward~\citep{wu2026}. In the model, the combined effect of the gas and magnetic pressure inside the plasmoids drives them to expand, while the total pressure in the open field falls off with increasing radius. The plasmoids are magnetically isolated island structures; hence, conductive heat flux cannot equalise their temperature with their surroundings. As a result, they undergo adiabatic cooling while expanding. The temperature can drop from 1.56 MK to 0.66 MK between $6.8~R_{\odot}-24~R_{\odot}$.  We found that for slightly different choices of the coronal heating profile (such as applying Equation (\ref{eq:heatingfn}) without the $1/r^2$ dependency), the stripping of flux from the helmet streamer can be even more dramatic \citep[see also][]{Suess96}. We return to discuss this aspect further in Section \ref{sec:discuss}. 
\subsection{Pseudostreamer dynamics}
\label{subsec:pseudo_dyn}
{Turning now to the evolution of the  Pseudostreamer, the most prominent feature is the presence of Interchange Magnetic Reconnection (IMR) near the apex of the pseudostreamer, {at $r\approx 1.75~R_{\odot}$ -- see Figure \ref{fig:dpseudo}.} 
The two closed magnetic arcades that form the base of the Pseudostreamer are in general asymmetric in size, as the radial magnetic flux is not symmetric across the photospheric polarity inversion line at the base of the pseudostreamer. In general, the closed loops from one half the streamer will be longer than the other and therefore receive larger total coronal heating. As is well-known from the so-called loop scaling laws, the coronal gas pressure increases with increasing loop length for fixed coronal heating rate \citep[e.g.][]{rosner1978}.  The imbalance in gas pressure causes the closed loops near the pseudostreamer apex, where the plasma beta is high, to expand and compress against the oppositely oriented open magnetic field.
This forms a current sheet, evident from the reversal of the radial component of ${\bf B}$ across this layer.
 As a result, IMR takes place. In Figure~\ref{fig:dpseudo}(a), a small-scale plasmoid is apparent at the upper end of the current sheet, confirming the presence of IMR in this magnetic configuration. 
 During this interchange reconnection, open and closed magnetic field lines reconnect with one another without any net opening or closing of flux. 
 After IMR occurs, part of the previously closed magnetic flux becomes open and exactly the same amount of open flux becomes closed. This distinguishes the reconnection taking place at the pseudostreamer from that at the Helmet streamer stalk, which involves primarily open-open or closed-closed reconnection, at least for this 2.5D model.
 As the IMR removes flux from one closed arcade and adds it to the other, the latter starts to expand, choking off and eventually reversing the orientation of the current sheet, inducing IMR on the opposite flank of the pseudostreamer, and the whole process repeats
 quasi-periodically. 
 In Figure.~\ref{fig:dpseudo}~(b)--(d) panels, different phases of the IMR can be observed. We observe this repeated reconnection process both during the quasi-steady and driving phase, although the properties are different. In the absence of the shear driving, IMR is solely induced by the interplay of local coronal heating, conduction and radiative cooling. In the driving phase, both local thermodynamics and shear driving contribute to exciting IMR, and the reconnection process becomes more dynamic, with the formation of a greater number of plasmoids.
 By driving the shear flow at the photosphere, magnetic free energy in the form of an out-of-plane ($\phi $ component) field builds up and propagates to the null region where it leads to current sheet formation. A part of this free energy is dissipated by the IMR process and in the process, plasmoids are ejected. It is important to note that even though the driving motions are constant in one direction, the reconnection alternates back and forth between the two closed flux systems. Since the driving speed is very slow compared to the coronal Alfvén speed, then in principle it should be possible for the system to achieve a steady state where the IMR is all in one direction and smooth. Instead, we find bursty reconnection that seems to overshoot, so that the IMR alternates back and forth.  
 
 Another key point is that, although there are quasi-periodic plasmoid ejections from the reconnection site, the plasmoids cannot be advected by the background solar wind for a large distance. The background open magnetic field of the solar wind interacts with the plasmoid's field, reconnects with one side of the plasmoid and they are quickly eroded away and disappear, as observed in many jet models \citep[e.g.][]{pariat2009}. The net result is to produce a large flux of nonlinear Alfvén waves that carry the energy into the heliosphere, and may well be the origin of the celebrated ``Switchbacks" that have been frequently observed by PSP \citep{Wyper2022}. We discuss the properties of these waves further in the following section.
 
 It is noteworthy that for exciting and sustaining reconnection through the tearing mode instability, a necessary condition is to reach beyond the critical Lundquist number, which is of the order of $10^4$~\citep{biskamp1986,Loureiro2007,Bhattacharjee2009,Gannouni2023}. In our model, we find a large presence of plasmoids in the reconnecting current sheets when the refinement level is increased from 5 to 7 levels, i.e., when the local grid spacing is reduced from 1 Mm to 0.25 Mm.
\section{Synthetic In-Situ observation at streamers}\label{sec:in-situ}
\begin{figure}
    \centering
    \includegraphics[width=\linewidth]{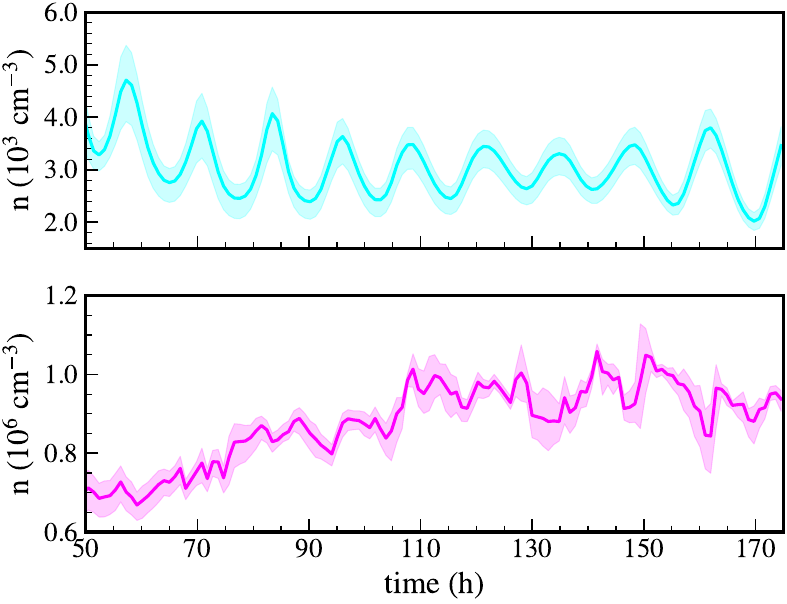}
    \caption{In-situ measurement of the electron number density from the Solar wind model: In panel (a), in situ electron number density is sampled over a line segment of length $0.1~R_\odot$ above the Helmet streamer stalk at $r\approx 16~R_\odot$ (as shown by the white arrow annotation in Fig.~\ref{fig:dhelmet}(a)). The cyan line plot shows the evolution of the average number density over sampled points and the background area represents the minimum and maximum value of density over the line segment. In panel (b), we follow the same procedure for the equal angular width arc as panel (a), which is $0.01~R_\odot$ line segment at $r\approx 2.5~R_\odot$, located near the Pseudostreamer stalk (see the arrow in Fig.~\ref{fig:dpseudo}(a))}.
\label{fig:n_prof}
\end{figure}
The hot, dense plasma released from the closed field regions by the reconnection processes described in the previous section should be transported out into the heliosphere, leading to fluctuations in observable quantities.
From our model, we calculate synthetic in situ measurements of the electron number density ($n$) above the stalk of both streamers. This is done by simply calculating $n$ at two sets of points covering the stalks. The in-situ observation points for the Helmet streamer are selected along a $0.1~R_\odot$ straight line, positioned at $r\approx 16~R_\odot$ (see the white arrow marker in  Figure~\ref{fig:dhelmet}(a)). We chose this location because Parker Solar Probe recently measured in-situ plasma quantities at a similar distance from the Sun. In Figure~\ref{fig:n_prof}(a), the variation of the electron number density at this location is shown. Both the spatially averaged profile (blue-solid curve) and the minimum to maximum number density range (blue shaded region) show a clear periodicity between 11.5--14.0 hours with varying amplitudes. 
This density oscillation over the Helmet streamer stalk is due to the quasi-periodic plasmoid ejections (see Section \ref{sec:dynamics}), which propagate away from the Sun and across the measurement points (Figure~\ref{fig:dhelmet}). During its  
8$^{\rm th}$ solar encounter, 
Parker Solar Probe was located at $r\approx 16~R_\odot$ from the Sun, and 
using quasi-thermal noise spectroscopy, the local electron density was measured at around ~$6\times10^3-7\times10^3$ cm$^{-3}$~\citep{kasper2021,Kruparova2023}. 
This result matches in order of magnitude with our simulated values at a similar distance. 

To measure the open-field density variation induced by the Pseudostreamer dynamics, we select in-situ observation points along an equal angular width arc as the Helmet streamer case above the Pseudostreamer stalk, which is $0.01~R_\odot$ arc length at $r\approx 2.5~R_\odot$ (see the white arrow marker in Figure.~\ref{fig:dpseudo}(a)). The averaged electron density over the selected points (dashed magenta line in Figure.~\ref{fig:n_prof}(b)) exhibits a fluctuating nature with time. 
This fluctuating temporal behaviour of the electron density profile can be described in terms of the induced IMR process at the Pseudostreamer. 
As we discuss in Subsection~\ref{subsec:pseudo_dyn}, dense plasma elements from the closed magnetic arcades of the Pseudostreamer can jump onto open field lines through the IMR mechanism. Then these dense plasma parcels propagate outward as plasmoids with the solar wind and contribute to a bursty electron density measurement. In the latest in-situ observations of PSP, fluctuating plasma characteristics have been well observed for several close encounters with the sun. Although the origin of the fluctuations is not completely understood, we believe several observed aspects can be described qualitatively by simple IMR models like our solar wind model.
\begin{figure}
    \centering
    \includegraphics[width=\linewidth]{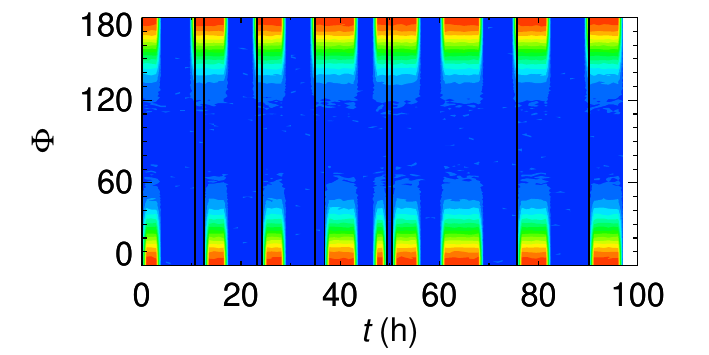}
    \caption{ A temporal map of the synthetic electron Strahl: it is computed at a point located above the Helmet streamer stalk at $r=16~R_{\odot}$. Here, $\Phi$ represents the relative angle between the magnetic field and the heat flux carried by the strahl.
    The RGB vertical segment corresponds to heat fluxes from the closed-magnetic field lines. A blue vertical patch indicates the absence of strahl, which is associated with the plasmoids. The narrow region between the black vertical lines represents the fluxes that propagate away from the sun.}
\label{fig:strahl}
\end{figure}

 To gain further insights in terms of the in-situ observables, we investigate a synthetic Strahl map from our model. Strahl is a measure of suprathermal electron heat flux that indicates the magnetic field orientation and the nature of magnetic connectivity~\citep{Feldman75,pilipp87}. It is routinely observed in situ by several space-based observation facilities, e.g., Wind, Helios, Parker Solar Probe, Solar Orbiter, etc. From our model, we compute the synthetic in-situ strahl map (Fig.~\ref{fig:strahl}) for a fixed point above the Helmet streamer stalk at $r=16~R_{\odot}$ by evaluating the magnetic field line connectivity for the field line passing through that point, as a function of time.

The map clearly shows the source of strahl electron from three different types of magnetic structures; (i) strahl from closed magnetic loop (denoted by RGB vertical sections) (ii) no strahl from magnetically isolated plasmoids (blue vertical strip) (iii) strahl from open flux, where the foot point is anchored at the negative polarity region of the chromosphere (narrow area between vertical black lines). In the strahl map, we find repeated occurrences of no strahl phases, due to the quasi-periodic plasmoid ejection at the HCS. 
Since the point we have selected for this calculation is to one side of the HCS, we note that when the field is open, it is always connected to negative polarity at the base (corresponding to a strahl signal around $\Phi=180$). 
We emphasise that our simulation lacks the complex physics to generate suprathermal electrons -- the map is a visualisation of the field line connectivity alone, but it reveals the periodic strahl variations that we would expect to find in observations for such a field evolution.
\newline
To explore the nature of the propagating disturbances in our simulations, 
we compute the Alfvénicity or normalized cross-helicity ($\sigma$). This quantity measures the correlation between velocity and magnetic field fluctuations following Alfvén wave properties~\citep{Barnes79,Matthaeus1982,Tu1995}. The definition of Alfvénicity is as follows, \newline $\sigma=<2{\boldsymbol{\delta}}\uu(t)\cdot{\boldsymbol{\delta}}v_{A}(t)>/<(\boldsymbol{\delta}\uu(t)^2+\boldsymbol{\delta}v_{A}(t)^2)>$, where $\boldsymbol{\delta}\uu(t)=\uu(t)-<\uu(t)>$,\\
$\boldsymbol{\delta}v_{A}(t)=(\bb(t)-<\bb(t)>)/\sqrt{4\pi<\rho(t)>}$ in cgs unit and $<~>$ represents a temporal average, chosen here to be over half an hour period. We calculate fluctuations in the velocity field $(\boldsymbol{\delta}\uu)$ and magnetic field $(\boldsymbol{\delta}\bb)$ with 1-minute cadence. In Figure~\ref{fig:sigma}, the Alfvénicity is shown at $r=16~R_{\odot}$ over a range of colatitudes, covering both the Helmet and Pseudostreamer. To remove any artifact of the imposed polar boundary conditions, we neglect regions close to the poles. 
We see that the Alfvénicity changes sign across the HCS, in general taking negative~(positive) values when the temporally averaged radial magnetic field is positive~(negative). These characteristics affirm that all of the fluctuations embedded in the solar wind are propagating away from the sun at $16~R_{\odot}$. 
A very high level of Alfvénicity $\sigma=-0.9$ is detected in the colatitude range of 110-120 degrees (Figure~\ref{fig:sigma}); this location corresponds to the Pseudostreamer in the lower atmosphere. This is one of the crucial results from our model that suggests Pseudostreamers can contribute to a strong level of Alfvénic tubulence by exciting Alfvénic waves during the IMR process. Recent observations from the Solar Orbiter reveal that a Pseudostreamer does excite a high level of Alfvénicity~\citep{Amicis25} in the slow solar wind. It is also intriguing that fluctuations that originate near the apex of the Pseutostreamer at $r\approx 2.2 R_{\odot}$ in our model can survive to a much longer distance at $16~R_{\odot}$ and show their signatures in terms of Alfvénicity measurement. Unlike the Pseudostreamer, the Helmet streamer does not exhibit any sign of high Alfvénic activity. Finally, it should be noted that the fine features of the Alfvénicity curve strongly depend on the dynamic background state and the time period over which fluctuations are computed, but the general nature of the curve remains the same.

\begin{figure}
    \centering
    \includegraphics[width=0.8\linewidth]{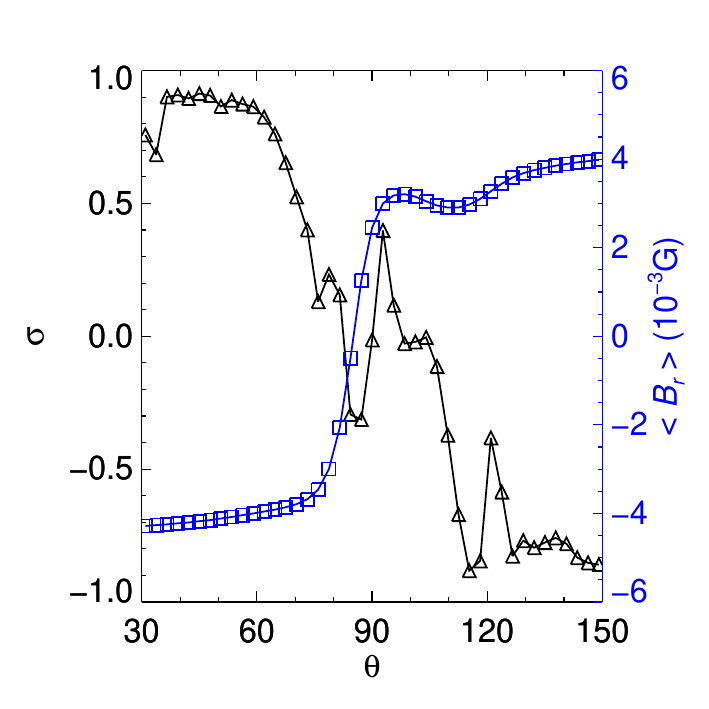}
    \caption{Variation of Alfvénicity measurement $(\sigma)$ as a function of colatitude is represented with the black curve at $r=16~R_{\odot}$. The radial component of the magnetic field is averaged over half an hour duration and shown by the blue curve.}
\label{fig:sigma}
\end{figure}
\section{Discussion}\label{sec:discuss}
In this paper our goal was to explore the interaction of closed and open magnetic field (and associated plasma) under generic conditions on the Sun. That is, we are interested not in the rare, large eruptions from the closed field, but rather in the quiescent behaviour driven by the continual sub-surface convective flows. 
In order to extend previous studies of interchange reconnection -- and in particular to enable closer connections to be made to the observations -- we have included in our model a chromospheric layer, as well as field-aligned thermal conduction, radiative losses, and empirical heating functions. 
With these physical effects included, we obtain a quasi-steady state with an outflowing wind in which the plasma in the closed-field region is both hotter (by a factor of 1.40--1.72) and denser (by a factor of $\approx$ 8) than in the adjacent open field, broadly consistent with observations. As such, when magnetic flux changes connectivity from closed to open during the simulations, the hot, dense plasma can expand into the cooler, more tenuous surroundings along the newly open flux tube.

Even in the absence of any boundary driving, both the helmet streamer and pseudostreamer show an ongoing, periodic dynamics. For the helmet streamer this is expected based on numerous previous studies, but for the pseudostreamer, we speculate that the dynamics are generated/maintained by the plasma thermodynamics in the closed field, and have thus not been observed before in simulations employing a simplified treatment of the thermodynamics. As such, it will be of interest in future to explore whether this behaviour survives when different heating functions are used, for different sizes of pseudostreamer, in three dimensions, etc.  For both the helmet streamer and pseudostreamer, the dynamics become more pronounced when boundary shearing is introduced to mimic the effects of supergranular driving.

At the helmet streamer, the interaction between closed and open flux takes the form of a ``breathing'' cycle in which the closed flux contracts and expands, and plasmoids are released along the heliospheric current sheet.  The period of this ``pinching off'' of plasmoids in our simulations is around 11.5--14.0 hours, and leads to a similar periodicity in density fluctuations in the vicinity of the heliospheric current sheet.
The imprint of these quasi-periodic plasmoid ejections is also well captured in the in-situ diagnostics. e.g., the Strahl map.
However, we note that a preliminary exploration indicates that the periodicity depends on the grid refinement that we permit along the heliospheric current sheet. This is in spite of the fact that the simulation never reaches maximum refinement in that region (when the maximum refinement level exceeds 5). Therefore, in a future study, we will make a systematic study where we vary the grid refinement level and strategy to seek a limiting value for this period.

By contrast to the helmet streamer, at the pseudostreamer, the plasma exchange between open and closed field must be mediated by interchange reconnection in which, in any finite time window, an equal amount of magnetic flux is opening on one side and closing on the other side of the pseudostreamer (no net opening or closing).
The pseudostreamer exhibits a back-and-forth motion, driving interchange reconnection alternately on its opposite flanks, even though in our simulation the driving is maintained in a constant direction. The resulting release of hot, dense plasma leads to density fluctuations in the open field that are significantly larger above the helmet streamer due to the persistence of the plasmoids there. An important signature of interchange magnetic reconnection that is implied by our simulations is the degree of Alfvénic fluctuations in the solar wind. We predict that a strong level of Alfvénicity/Cross Helicity would be detected over the region of Pseudostreamers, originating from the reconnection site near the apex of the same streamer.

Our model demonstrates that plasma is continually being released into the heliosphere from both streamers and pseudostreamers. As a result, across the S-Web we should expect to see plasma with mixed properties in terms of the magnetic connectivity at their origins in the low corona. Fluctuations in plasma properties above helmet streamers should by larger in magnitude, because plasmoids that form during the reconnection process can be carried out into the heliosphere intact. 
Even in 3D -- where these plasmoids are not strictly magnetically closed -- the magnetic flux rope structure of the plasmoid will at least inhibit mixing between the newly released plasma and the ambient wind. 
In contrast, plasma released from the pseudostreamer is not magnetically confined, since due to the field orientation, the (much smaller) plasmoids formed during interchange reconnection there are quickly eroded by reconnection between the plasmoid field and ambient open field. As such, the plasma released from the closed field region beneath the pseudostreamer quickly mixes with the adjacent open-field plasma. Despite the decay in plasmoid structures, the signatures of the interchange reconnection process are carried to several solar radii distances by the propagating Alfvénic waves.

There are a number of limitations to the present study, that we plan to overcome in future works. (i) The most obvious is the 2.5D nature of the simulations. 3D offers a much broader range of magnetic -- and therefore plasma -- dynamics, particularly in terms of the formation of new, open flux \citep[e.g.,][]{aslanyan2022}. It may well be that in 3D, more turbulence is generated in situ in the open field rather than primarily at the open-closed boundary. This may affect the level of Alfvénicity measured in the wind. In addition, as mentioned above, plasmoids are not magnetically closed structures in 3D, with important implications for plasma transport to larger radii. Due to this change for the nature of magnetic connectivity in 3D, the strahl from these plasmoids would be categorized as a contribution from the open magnetic field. 
(ii) Our solar wind is rather slow compared to the observed value: faster winds can be obtained by, for example, adding an additional Alfv\'en wave term to the momentum equation. (iii) As mentioned above, with the present grid refinement strategy we could not find a definitive, resolution-independent periodicity for the fluctuations induced. In future we will explore refining on, for example, the current density. (iv) Tracking plasma parcels released from the closed field into the open field would permit a more detailed comparison with observations, and will also be implemented in the future. 
\begin{acknowledgments}
S.D. and D.P. gratefully acknowledge support through an Australian Research Council Discovery Project (DP210100709). This research was supported by the Australian Government's National Collaborative
Research Infrastructure Strategy (NCRIS), with access to computational resources provided by the National Computational Infrastructure (NCI) through the National Computational Merit Allocation Scheme (NCMAS). S.D. and D.P. also acknowledge the generous computing grant provided by the AAL Supercomputer Time Allocation Committee (ASTAC). SKA acknowledges support from a NASA LWS grant to the University of Michigan. The authors acknowledge the helpful discussions with Craig Johnston and Cooper Downs. S.D. is grateful to Chun Xia, Yang Guo, and Rony Keppens for extensive discussions and suggestions regarding various numerical and physical aspects of the solar wind modelling.
\end{acknowledgments}
\software{MPI-AMRVAC 3.2~\citep{keppens2023}, Paraview~\citep{paraview2011}}




\bibliography{streamers}{}

@ARTICLE{abbo2016,
       author = {{Abbo}, L. and {Ofman}, L. and {Antiochos}, S.~K. and {Hansteen}, V.~H. and {Harra}, L. and {Ko}, Y.-K. and {Lapenta}, G. and {Li}, B. and {Riley}, P. and {Strachan}, L. and {von Steiger}, R. and {Wang}, Y.-M.},
        title = "{Slow Solar Wind: Observations and Modeling}",
      journal = {\ssr},
         year = 2016,
        month = nov,
       volume = {201},
       number = {1-4},
        pages = {55-108},
          doi = {10.1007/s11214-016-0264-1}
}

@ARTICLE{allred2015,
       author = {{Allred}, J.~C. and {MacNeice}, P.~J.},
        title = "{An MHD code for the study of magnetic structures in the solar wind}",
      journal = {Computational Science and Discovery},
         year = 2015,
        month = jan,
       volume = {8},
       number = {1},
          eid = {015002},
        pages = {015002},
          doi = {10.1088/1749-4680/8/1/015002},
       adsurl = {https://ui.adsabs.harvard.edu/abs/2015CS&D....8a5002A}
}

@ARTICLE{antiochos2011,
       author = {{Antiochos}, S.~K. and {Miki{\'c}}, Z. and {Titov}, V.~S. and
         {Lionello}, R. and {Linker}, J.~A.},
        title = "{A Model for the Sources of the Slow Solar Wind}",
      journal = {Astrophys.~J.},
         year = "2011",
        month = "Apr",
       volume = {731},
       number = {2},
          eid = {112},
        pages = {112},
          doi = {10.1088/0004-637X/731/2/112}
}

@ARTICLE{aslanyan2021,
       author = {{Aslanyan}, V. and {Pontin}, D.~I. and {Wyper}, P.~F. and {Scott}, R.~B. and {Antiochos}, S.~K. and {DeVore}, C.~R.},
        title = "{Effects of Pseudostreamer Boundary Dynamics on Heliospheric Field and Wind}",
      journal = {Astrophys.~J.},
         year = 2021,
        month = mar,
       volume = {909},
       number = {1},
          eid = {10},
        pages = {10},
          doi = {10.3847/1538-4357/abd6e6}
}

@ARTICLE{aslanyan2022,
       author = {{Aslanyan}, V. and {Pontin}, D.~I. and {Higginson}, A.~K. and {Wyper}, P.~F. and {Scott}, R.~B. and {Antiochos}, S.~K.},
        title = "{The Dynamic Coupling of Streamers and Pseudostreamers to the Heliosphere}",
      journal = {Astrophys.~J.},
         year = 2022,
        month = apr,
       volume = {929},
       number = {2},
          eid = {185},
        pages = {185},
          doi = {10.3847/1538-4357/ac5d5b}
}

@ARTICLE{Amicis25,
       author = {{D'Amicis}, R. and {Velli}, M. and {Panasenco}, O. and {Sorriso-Valvo}, L. and {Perrone}, D. and {Benella}, S. and {De Marco}, R. and {Bruno}, R. and {Wang}, Y.-M. and {R{\'e}ville}, V. and {Baker}, D. and {Matteini}, L. and {Yardley}, S. and {Settino}, A. and {Sioulas}, N. and {Alterman}, B. and {Tenerani}, A. and {Raines}, J. and {Holmes}, J. and {Buchlin}, E. and {Verdini}, A. and {Demoulin}, P. and {van Driel-Gesztelyi}, L. and {Telloni}, D. and {Consolini}, G. and {Marcucci}, M.~F. and {Stangalini}, M. and {Marino}, R. and {Fortunato}, V. and {Mele}, G. and {Monti}, F. and {Owen}, C.~J. and {Louarn}, P. and {Livi}, S.},
        title = "{On Alfv{\'e}nic turbulence of solar wind streams observed by Solar Orbiter during March 2022 perihelion and their source regions}",
      journal = {\aap},
         year = 2025,
        month = jan,
       volume = {693},
          eid = {A243},
        pages = {A243},
          doi = {10.1051/0004-6361/202451686},
       adsurl = {https://ui.adsabs.harvard.edu/abs/2025A&A...693A.243D}
}

@ARTICLE{Arge2024,
       author = {{Arge}, C. Nick and {Leisner}, Andrew and {Antiochos}, Spiro K. and {Wallace}, Samantha and {Henney}, Carl J.},
        title = "{Proposed Resolution to the Solar Open Magnetic Flux Problem}",
      journal = {\apj},
         year = 2024,
        month = apr,
       volume = {964},
       number = {2},
          eid = {115},
        pages = {115},
          doi = {10.3847/1538-4357/ad20e2},
archivePrefix = {arXiv},
       eprint = {2304.07649},
 primaryClass = {astro-ph.SR},
       adsurl = {https://ui.adsabs.harvard.edu/abs/2024ApJ...964..115A}
}

@INCOLLECTION{Barnes79,
       author = {{Barnes}, A.},
        title = "{Hydromagnetic waves and turbulence in the solar wind}",
    booktitle = {Solar System Plasma Physics},
         year = 1979,
       editor = {{Parker}, E.~N. and {Kennel}, C.~F. and {Lanzerotti}, L.~J.},
       volume = {1},
        pages = {249-319},
       adsurl = {https://ui.adsabs.harvard.edu/abs/1979sswp.book..249B}
}

@ARTICLE{biskamp1986,
       author = {{Biskamp}, D.},
        title = "{Magnetic reconnection via current sheets}",
      journal = {Physics of Fluids},
         year = 1986,
        month = may,
       volume = {29},
       number = {5},
        pages = {1520-1531},
          doi = {10.1063/1.865670},
       adsurl = {https://ui.adsabs.harvard.edu/abs/1986PhFl...29.1520B}
}

@article{borovsky2016,
author = {Borovsky, Joseph E.},
title = {The plasma structure of coronal hole solar wind: Origins and evolution},
journal = {Journal of Geophysical Research: Space Physics},
volume = {121},
number = {6},
pages = {5055-5087},
doi = {https://doi.org/10.1002/2016JA022686},
url = {https://agupubs.onlinelibrary.wiley.com/doi/abs/10.1002/2016JA022686},
eprint = {https://agupubs.onlinelibrary.wiley.com/doi/pdf/10.1002/2016JA022686},
year = {2016}
}

@ARTICLE{Bhattacharjee2009,
       author = {{Bhattacharjee}, A. and {Huang}, Yi-Min and {Yang}, H. and {Rogers}, B.},
        title = "{Fast reconnection in high-Lundquist-number plasmas due to the plasmoid Instability}",
      journal = {Physics of Plasmas},
         year = 2009,
        month = nov,
       volume = {16},
       number = {11},
          eid = {112102},
        pages = {112102},
          doi = {10.1063/1.3264103},
archivePrefix = {arXiv},
       eprint = {0906.5599},
 primaryClass = {physics.plasm-ph},
       adsurl = {https://ui.adsabs.harvard.edu/abs/2009PhPl...16k2102B}
}

@ARTICLE{brooks2015,
       author = {{Brooks}, David H. and {Ugarte-Urra}, Ignacio and {Warren}, Harry P.},
        title = "{Full-Sun observations for identifying the source of the slow solar wind}",
      journal = {Nature Communications},
         year = 2015,
        month = jan,
       volume = {6},
          eid = {5947},
        pages = {5947},
          doi = {10.1038/ncomms6947},
archivePrefix = {arXiv},
       eprint = {1605.09514},
 primaryClass = {astro-ph.SR},
       adsurl = {https://ui.adsabs.harvard.edu/abs/2015NatCo...6.5947B}
}

@ARTICLE{cranmer2019,
       author = {{Cranmer}, Steven R. and {Winebarger}, Amy R.},
        title = "{The Properties of the Solar Corona and Its Connection to the Solar Wind}",
      journal = {\araa},
         year = 2019,
        month = aug,
       volume = {57},
        pages = {157-187},
          doi = {10.1146/annurev-astro-091918-104416},
archivePrefix = {arXiv},
       eprint = {1811.00461},
 primaryClass = {astro-ph.SR},
       adsurl = {https://ui.adsabs.harvard.edu/abs/2019ARA&A..57..157C}
}

@ARTICLE{dere2009,
       author = {{Dere}, K.~P. and {Landi}, E. and {Young}, P.~R. and {Del Zanna}, G. and {Landini}, M. and {Mason}, H.~E.},
        title = "{CHIANTI - an atomic database for emission lines. IX. Ionization rates, recombination rates, ionization equilibria for the elements hydrogen through zinc and updated atomic data}",
      journal = {\aap},
         year = 2009,
        month = may,
       volume = {498},
       number = {3},
        pages = {915-929},
          doi = {10.1051/0004-6361/200911712},
       adsurl = {https://ui.adsabs.harvard.edu/abs/2009A&A...498..915D}
}

@article{downs2010,
doi = {10.1088/0004-637X/712/2/1219},
url = {https://doi.org/10.1088/0004-637X/712/2/1219},
year = {2010},
month = {mar},
publisher = {The American Astronomical Society},
volume = {712},
number = {2},
pages = {1219},
author = {Downs, Cooper and Roussev, Ilia I. and van der Holst, Bart and Lugaz, Noé and Sokolov, Igor V. and Gombosi, Tamas I.},
title = {TOWARD A REALISTIC THERMODYNAMIC MAGNETOHYDRODYNAMIC MODEL OF THE GLOBAL SOLAR CORONA},
journal = {The Astrophysical Journal}
}

@ARTICLE{edmondson2010,
       author = {{Edmondson}, J.~K. and {Antiochos}, S.~K. and {DeVore}, C.~R. and {Lynch}, B.~J. and {Zurbuchen}, T.~H.},
        title = "{Interchange Reconnection and Coronal Hole Dynamics}",
      journal = {Astrophys.~J.},
         year = 2010,
        month = may,
       volume = {714},
       number = {1},
        pages = {517-531},
          doi = {10.1088/0004-637X/714/1/517},
       adsurl = {https://ui.adsabs.harvard.edu/abs/2010ApJ...714..517E}
}

@ARTICLE{Feldman75,
       author = {{Feldman}, W.~C. and {Asbridge}, J.~R. and {Bame}, S.~J. and {Montgomery}, M.~D. and {Gary}, S.~P.},
        title = "{Solar wind electrons}",
      journal = {\jgr},
         year = 1975,
        month = nov,
       volume = {80},
       number = {31},
        pages = {4181},
          doi = {10.1029/JA080i031p04181},
       adsurl = {https://ui.adsabs.harvard.edu/abs/1975JGR....80.4181F}
}

@ARTICLE{Fan2017,
       author = {{Fan}, Yuhong},
        title = "{MHD Simulations of the Eruption of Coronal Flux Ropes under Coronal Streamers}",
      journal = {\apj},
         year = 2017,
        month = jul,
       volume = {844},
       number = {1},
          eid = {26},
        pages = {26},
          doi = {10.3847/1538-4357/aa7a56},
archivePrefix = {arXiv},
       eprint = {1706.06076},
 primaryClass = {astro-ph.SR},
       adsurl = {https://ui.adsabs.harvard.edu/abs/2017ApJ...844...26F}
}

@ARTICLE{Gannouni2023,
       author = {{Gannouni}, Bahaeddine and {R{\'e}ville}, Victor and {Rouillard}, Alexis P.},
        title = "{Modeling the Formation and Evolution of Solar Wind Microstreams: From Coronal Plumes to Propagating Alfv{\'e}nic Velocity Spikes}",
      journal = {\apj},
         year = 2023,
        month = dec,
       volume = {958},
       number = {2},
          eid = {110},
        pages = {110},
          doi = {10.3847/1538-4357/acfef3},
archivePrefix = {arXiv},
       eprint = {2307.02210},
 primaryClass = {astro-ph.SR},
       adsurl = {https://ui.adsabs.harvard.edu/abs/2023ApJ...958..110G}
}

@ARTICLE{Gottlieb1998,
       author = {{Gottlieb}, S. and {Shu}, C.~W.},
        title = "{Total variation diminishing Runge-Kutta schemes}",
      journal = {Mathematics of Computation},
         year = 1998,
        month = jan,
       volume = {67},
       number = {221},
        pages = {73-85},
          doi = {10.1090/S0025-5718-98-00913-2},
       adsurl = {https://ui.adsabs.harvard.edu/abs/1998MaCom..67...73G}
}

@ARTICLE{eastwood2026,
       author = {{Eastwood}, J.~P. and {Phan}, T.~D. and {Drake}, J.~F. and {Shay}, M.~A. and {{\O}ieroset}, M. and {Fargette}, N. and {Waters}, C.~L. and {Lewis}, H.~C. and {Badman}, S.~T. and {Stevens}, M.~L. and {Halekas}, J. and {Bale}, S.~D.},
        title = "{Magnetic Reconnection Energy Fluxes in the Near-Sun Heliospheric Current Sheet as Observed by Parker Solar Probe}",
      journal = {\apj},
         year = 2026,
        month = jan,
       volume = {996},
       number = {2},
          eid = {140},
        pages = {140},
          doi = {10.3847/1538-4357/ae1fe1},
       adsurl = {https://ui.adsabs.harvard.edu/abs/2026ApJ...996..140E}
}

@article{iijima2023,
doi = {10.3847/2041-8213/acdde0},
url = {https://doi.org/10.3847/2041-8213/acdde0},
year = {2023},
month = {jul},
publisher = {The American Astronomical Society},
volume = {951},
number = {2},
pages = {L47},
author = {Iijima, Haruhisa and Matsumoto, Takuma and Hotta, Hideyuki and Imada, Shinsuke},
title = {A Comprehensive Simulation of Solar Wind Formation from the Solar Interior: Significant Cross-field Energy Transport by Interchange Reconnection near the Sun},
journal = {The Astrophysical Journal Letters}
}

@ARTICLE{keppens2003,
       author = {{Keppens}, R. and {Nool}, M. and {T{\'o}th}, G. and {Goedbloed}, J.~P.},
        title = "{Adaptive Mesh Refinement for conservative systems: multi-dimensional efficiency evaluation}",
      journal = {Computer Physics Communications},
         year = 2003,
        month = jul,
       volume = {153},
       number = {3},
        pages = {317-339},
          doi = {10.1016/S0010-4655(03)00139-5},
archivePrefix = {arXiv},
       eprint = {astro-ph/0403124},
 primaryClass = {astro-ph},
       adsurl = {https://ui.adsabs.harvard.edu/abs/2003CoPhC.153..317K}
}

@ARTICLE{keppens2020,
       author = {{Keppens}, Rony and {Teunissen}, Jannis and {Xia}, Chun and {Porth}, Oliver},
        title = "{MPI-AMRVAC: a parallel, grid-adaptive PDE toolkit}",
      journal = {arXiv e-prints},
         year = 2020,
        month = apr,
          eid = {arXiv:2004.03275},
        pages = {arXiv:2004.03275},
          doi = {10.48550/arXiv.2004.03275},
archivePrefix = {arXiv},
       eprint = {2004.03275},
 primaryClass = {astro-ph.IM},
       adsurl = {https://ui.adsabs.harvard.edu/abs/2020arXiv200403275K}
}

@ARTICLE{kasper2021,
       author = {{Kasper}, J.~C. and {Klein}, K.~G. and {Lichko}, E. and {Huang}, Jia and {Chen}, C.~H.~K. and {Badman}, S.~T. and {Bonnell}, J. and {Whittlesey}, P.~L. and {Livi}, R. and {Larson}, D. and {Pulupa}, M. and {Rahmati}, A. and {Stansby}, D. and {Korreck}, K.~E. and {Stevens}, M. and {Case}, A.~W. and {Bale}, S.~D. and {Maksimovic}, M. and {Moncuquet}, M. and {Goetz}, K. and {Halekas}, J.~S. and {Malaspina}, D. and {Raouafi}, Nour E. and {Szabo}, A. and {MacDowall}, R. and {Velli}, Marco and {Dudok de Wit}, Thierry and {Zank}, G.~P.},
        title = "{Parker Solar Probe Enters the Magnetically Dominated Solar Corona}",
      journal = {\prl},
         year = 2021,
        month = dec,
       volume = {127},
       number = {25},
          eid = {255101},
        pages = {255101},
          doi = {10.1103/PhysRevLett.127.255101},
       adsurl = {https://ui.adsabs.harvard.edu/abs/2021PhRvL.127y5101K}
}

@ARTICLE{Kruparova2023,
       author = {{Kruparova}, Oksana and {Krupar}, Vratislav and {Szabo}, Adam and {Pulupa}, Marc and {Bale}, Stuart D.},
        title = "{Quasi-thermal Noise Spectroscopy Analysis of Parker Solar Probe Data: Improved Electron Density Model for Solar Wind}",
      journal = {\apj},
         year = 2023,
        month = nov,
       volume = {957},
       number = {1},
          eid = {13},
        pages = {13},
          doi = {10.3847/1538-4357/acf572},
       adsurl = {https://ui.adsabs.harvard.edu/abs/2023ApJ...957...13K}
}

@ARTICLE{Lohner1987,
       author = {{Lohner}, R.},
        title = "{An adaptive finite element scheme for transient problems in CFD}",
      journal = {Computer Methods in Applied Mechanics and Engineering},
         year = 1987,
        month = apr,
       volume = {61},
       number = {3},
        pages = {323-338},
          doi = {10.1016/0045-7825(87)90098-3},
       adsurl = {https://ui.adsabs.harvard.edu/abs/1987CMAME..61..323L}
}

@BOOK{Lamers1999,
       author = {{Lamers}, Henny J.~G.~L.~M. and {Cassinelli}, Joseph P.},
        title = "{Introduction to Stellar Winds}",
         year = 1999,
       adsurl = {https://ui.adsabs.harvard.edu/abs/1999isw..book.....L}
}

@ARTICLE{Loureiro2007,
       author = {{Loureiro}, N.~F. and {Schekochihin}, A.~A. and {Cowley}, S.~C.},
        title = "{Instability of current sheets and formation of plasmoid chains}",
      journal = {Physics of Plasmas},
         year = 2007,
        month = oct,
       volume = {14},
       number = {10},
        pages = {100703-100703},
          doi = {10.1063/1.2783986},
archivePrefix = {arXiv},
       eprint = {astro-ph/0703631},
 primaryClass = {astro-ph},
       adsurl = {https://ui.adsabs.harvard.edu/abs/2007PhPl...14j0703L}
}

@ARTICLE{Matthaeus1982,
       author = {{Matthaeus}, W.~H. and {Goldstein}, M.~L.},
        title = "{Measurement of the rugged invariants of magnetohydrodynamic turbulence in the solar wind}",
      journal = {\jgr},
         year = 1982,
        month = aug,
       volume = {87},
       number = {A8},
        pages = {6011-6028},
          doi = {10.1029/JA087iA08p06011},
       adsurl = {https://ui.adsabs.harvard.edu/abs/1982JGR....87.6011M}
}

@ARTICLE{matsumoto2021,
       author = {{Matsumoto}, Takuma},
        title = "{Full compressible 3D MHD simulation of solar wind}",
      journal = {\mnras},
         year = 2021,
        month = jan,
       volume = {500},
       number = {4},
        pages = {4779-4787},
          doi = {10.1093/mnras/staa3533},
archivePrefix = {arXiv},
       eprint = {2009.03770},
 primaryClass = {astro-ph.SR},
       adsurl = {https://ui.adsabs.harvard.edu/abs/2021MNRAS.500.4779M}
}

@ARTICLE{parker1958,
       author = {{Parker}, E.~N.},
        title = "{Dynamics of the Interplanetary Gas and Magnetic Fields.}",
      journal = {\apj},
         year = 1958,
        month = nov,
       volume = {128},
        pages = {664},
          doi = {10.1086/146579},
       adsurl = {https://ui.adsabs.harvard.edu/abs/1958ApJ...128..664P}
}

@ARTICLE{pilipp87,
       author = {{Pilipp}, W.~G. and {Miggenrieder}, H. and {M{\"u}hlha{\"u}ser}, K.-H. and {Rosenbauer}, H. and {Schwenn}, R. and {Neubauer}, F.~M.},
        title = "{Variations of electron distribution functions in the solar wind}",
      journal = {\jgr},
         year = 1987,
        month = feb,
       volume = {92},
       number = {A2},
        pages = {1103-1118},
          doi = {10.1029/JA092iA02p01103},
       adsurl = {https://ui.adsabs.harvard.edu/abs/1987JGR....92.1103P}
}

@ARTICLE{porth2014,
       author = {{Porth}, O. and {Xia}, C. and {Hendrix}, T. and {Moschou}, S.~P. and {Keppens}, R.},
        title = "{MPI-AMRVAC for Solar and Astrophysics}",
      journal = {\apjs},
         year = 2014,
        month = sep,
       volume = {214},
       number = {1},
          eid = {4},
        pages = {4},
          doi = {10.1088/0067-0049/214/1/4},
archivePrefix = {arXiv},
       eprint = {1407.2052},
 primaryClass = {astro-ph.IM},
       adsurl = {https://ui.adsabs.harvard.edu/abs/2014ApJS..214....4P}
}

@ARTICLE{rosner1978,
       author = {{Rosner}, R. and {Tucker}, W.~H. and {Vaiana}, G.~S.},
        title = "{Dynamics of the quiescent solar corona.}",
      journal = {\apj},
         year = 1978,
        month = mar,
       volume = {220},
        pages = {643-645},
          doi = {10.1086/155949},
       adsurl = {https://ui.adsabs.harvard.edu/abs/1978ApJ...220..643R}
}

@ARTICLE{Tu1995,
       author = {{Tu}, C.-Y. and {Marsch}, E.},
        title = "{Magnetohydrodynamic Structures Waves and Turbulence in the Solar Wind - Observations and Theories}",
      journal = {\ssr},
         year = 1995,
        month = jul,
       volume = {73},
       number = {1-2},
        pages = {1-210},
          doi = {10.1007/BF00748891},
       adsurl = {https://ui.adsabs.harvard.edu/abs/1995SSRv...73....1T}
}

@ARTICLE{talpeanu2022,
       author = {{Talpeanu}, D. -C. and {Poedts}, S. and {D'Huys}, E. and {Mierla}, M.},
        title = "{Study of the propagation, in situ signatures, and geoeffectiveness of shear-induced coronal mass ejections in different solar winds}",
      journal = {\aap},
         year = 2022,
        month = feb,
       volume = {658},
          eid = {A56},
        pages = {A56},
          doi = {10.1051/0004-6361/202141977},
archivePrefix = {arXiv},
       eprint = {2111.14909},
 primaryClass = {astro-ph.SR},
       adsurl = {https://ui.adsabs.harvard.edu/abs/2022A&A...658A..56T}
}

@ARTICLE{fisk1998,
       author = {{Fisk}, L.~A. and {Schwadron}, N.~A. and {Zurbuchen}, T.~H.},
        title = "{On the Slow Solar Wind}",
      journal = {\ssr},
         year = 1998,
        month = jul,
       volume = {86},
        pages = {51-60},
          doi = {10.1023/A:1005015527146},
       adsurl = {https://ui.adsabs.harvard.edu/abs/1998SSRv...86...51F}
}

@ARTICLE{Tanaka1994,
       author = {{Tanaka}, T.},
        title = "{Finite Volume TVD Scheme on an Unstructured Grid System for Three-Dimensional MHD Simulation of Inhomogeneous Systems Including Strong Background Potential Fields}",
      journal = {Journal of Computational Physics},
         year = 1994,
        month = apr,
       volume = {111},
       number = {2},
        pages = {381-389},
          doi = {10.1006/jcph.1994.1071},
       adsurl = {https://ui.adsabs.harvard.edu/abs/1994JCoPh.111..381T}
}

@ARTICLE{higginson2017b,
       author = {{Higginson}, A.~K. and {Antiochos}, S.~K. and {DeVore}, C.~R. and
         {Wyper}, P.~F. and {Zurbuchen}, T.~H.},
        title = "{Formation of Heliospheric Arcs of Slow Solar Wind}",
      journal = {Astrophys.~J.~Lett.},
         year = "2017",
        month = "May",
       volume = {840},
       number = {1},
          eid = {L10},
        pages = {L10},
          doi = {10.3847/2041-8213/aa6d72}
}

@ARTICLE{higginson2018,
       author = {{Higginson}, A.~K. and {Lynch}, B.~J.},
        title = "{Structured Slow Solar Wind Variability: Streamer-blob Flux Ropes and Torsional Alfv{\'e}n Waves}",
      journal = {\apj},
         year = 2018,
        month = may,
       volume = {859},
       number = {1},
          eid = {6},
        pages = {6},
          doi = {10.3847/1538-4357/aabc08},
archivePrefix = {arXiv},
       eprint = {1710.00106},
 primaryClass = {astro-ph.SR},
       adsurl = {https://ui.adsabs.harvard.edu/abs/2018ApJ...859....6H}
}

@ARTICLE{jacobs2005,
       author = {{Jacobs}, C. and {Poedts}, S. and {Van der Holst}, B. and {Chan{\'e}}, E.},
        title = "{On the effect of the background wind on the evolution  of interplanetary shock waves}",
      journal = {\aap},
         year = 2005,
        month = feb,
       volume = {430},
        pages = {1099-1107},
          doi = {10.1051/0004-6361:20041676},
       adsurl = {https://ui.adsabs.harvard.edu/abs/2005A&A...430.1099J}
}

@ARTICLE{jhonston2019,
       author = {{Johnston}, C.~D. and {Bradshaw}, S.~J.},
        title = "{A Fast and Accurate Method to Capture the Solar Corona/Transition Region Enthalpy Exchange}",
      journal = {\apjl},
         year = 2019,
        month = mar,
       volume = {873},
       number = {2},
          eid = {L22},
        pages = {L22},
          doi = {10.3847/2041-8213/ab0c1f},
archivePrefix = {arXiv},
       eprint = {1903.01132},
 primaryClass = {astro-ph.SR},
       adsurl = {https://ui.adsabs.harvard.edu/abs/2019ApJ...873L..22J}
}

@ARTICLE{jhonston2020,
       author = {{Johnston}, C.~D. and {Cargill}, P.~J. and {Hood}, A.~W. and {De Moortel}, I. and {Bradshaw}, S.~J. and {Vaseekar}, A.~C.},
        title = "{Modelling the solar transition region using an adaptive conduction method}",
      journal = {\aap},
         year = 2020,
        month = mar,
       volume = {635},
          eid = {A168},
        pages = {A168},
          doi = {10.1051/0004-6361/201936979},
archivePrefix = {arXiv},
       eprint = {2002.01887},
 primaryClass = {astro-ph.SR},
       adsurl = {https://ui.adsabs.harvard.edu/abs/2020A&A...635A.168J}
}

@ARTICLE{keppens2023,
       author = {{Keppens}, R. and {Popescu Braileanu}, B. and {Zhou}, Y. and {Ruan}, W. and {Xia}, C. and {Guo}, Y. and {Claes}, N. and {Bacchini}, F.},
        title = "{MPI-AMRVAC 3.0: Updates to an open-source simulation framework}",
      journal = {\aap},
         year = 2023,
        month = may,
       volume = {673},
          eid = {A66},
        pages = {A66},
          doi = {10.1051/0004-6361/202245359},
archivePrefix = {arXiv},
       eprint = {2303.03026},
 primaryClass = {astro-ph.IM},
       adsurl = {https://ui.adsabs.harvard.edu/abs/2023A&A...673A..66K}
}

@ARTICLE{lionello2009s,
       author = {{Lionello}, Roberto and {Linker}, Jon A. and {Miki{\'c}}, Zoran},
        title = "{Multispectral Emission of the Sun During the First Whole Sun Month: Magnetohydrodynamic Simulations}",
      journal = {\apj},
         year = 2009,
        month = jan,
       volume = {690},
       number = {1},
        pages = {902-912},
          doi = {10.1088/0004-637X/690/1/902},
       adsurl = {https://ui.adsabs.harvard.edu/abs/2009ApJ...690..902L}
}

@ARTICLE{mccomas2008,
       author = {{McComas}, D.~J. and {Ebert}, R.~W. and {Elliott}, H.~A. and {Goldstein}, B.~E. and {Gosling}, J.~T. and {Schwadron}, N.~A. and {Skoug}, R.~M.},
        title = "{Weaker solar wind from the polar coronal holes and the whole Sun}",
      journal = {\grl},
         year = 2008,
        month = sep,
       volume = {35},
       number = {18},
          eid = {L18103},
        pages = {L18103},
          doi = {10.1029/2008GL034896}
}

@ARTICLE{Maity2024,
       author = {{Sankar Maity}, Samriddhi and {Chatterjee}, Piyali and {Sarkar}, Ranadeep and {Mytheen}, Ijas S.},
        title = "{Evolution of reconnection flux during eruption of magnetic flux ropes}",
      journal = {arXiv e-prints},
         year = 2024,
        month = jul,
          eid = {arXiv:2407.18188},
        pages = {arXiv:2407.18188},
          doi = {10.48550/arXiv.2407.18188},
archivePrefix = {arXiv},
       eprint = {2407.18188},
 primaryClass = {astro-ph.SR},
       adsurl = {https://ui.adsabs.harvard.edu/abs/2024arXiv240718188S}
}

@ARTICLE{neugebauer1962,
       author = {{Neugebauer}, Marcia and {Snyder}, Conway W.},
        title = "{Solar Plasma Experiment}",
      journal = {Science},
         year = 1962,
        month = dec,
       volume = {138},
       number = {3545},
        pages = {1095-1097},
          doi = {10.1126/science.138.3545.1095-a},
       adsurl = {https://ui.adsabs.harvard.edu/abs/1962Sci...138.1095N}
}

@article{owens2013,
author = {Owens, M. J. and Crooker, N. U. and Lockwood, M.},
title = {Solar cycle evolution of dipolar and pseudostreamer belts and their relation to the slow solar wind},
journal = {Journal of Geophysical Research: Space Physics},
volume = {119},
number = {1},
pages = {36-46},
doi = {https://doi.org/10.1002/2013JA019412},
url = {https://agupubs.onlinelibrary.wiley.com/doi/abs/10.1002/2013JA019412},
eprint = {https://agupubs.onlinelibrary.wiley.com/doi/pdf/10.1002/2013JA019412},
year = {2014}
}

@ARTICLE{pariat2009,
       author = {{Pariat}, E. and {Antiochos}, S.~K. and {DeVore}, C.~R.},
        title = "{A Model for Solar Polar Jets}",
      journal = {\apj},
         year = 2009,
        month = jan,
       volume = {691},
       number = {1},
        pages = {61-74},
          doi = {10.1088/0004-637X/691/1/61},
       adsurl = {https://ui.adsabs.harvard.edu/abs/2009ApJ...691...61P}
}

@ARTICLE{pontin2013,
       author = {{Pontin}, D.~I. and {Priest}, E.~R. and {Galsgaard}, K.},
        title = "{On the Nature of Reconnection at a Solar Coronal Null Point above a Separatrix Dome}",
      journal = {Astrophys.~J.},
         year = 2013,
        month = sep,
       volume = {774},
       number = {2},
          eid = {154},
        pages = {154},
          doi = {10.1088/0004-637X/774/2/154}
}

@software{paraview2011,
       author = {{Numerous}},
        title = "{ParaView: Data Analysis and Visualization Application}",
 howpublished = {Astrophysics Source Code Library, record ascl:1103.014},
         year = 2011,
        month = mar,
          eid = {ascl:1103.014},
archivePrefix = {ascl},
       eprint = {1103.014},
       adsurl = {https://ui.adsabs.harvard.edu/abs/2011ascl.soft03014N}
}

@ARTICLE{powell1999,
       author = {{Powell}, Kenneth G. and {Roe}, Philip L. and {Linde}, Timur J. and {Gombosi}, Tamas I. and {De Zeeuw}, Darren L.},
        title = "{A Solution-Adaptive Upwind Scheme for Ideal Magnetohydrodynamics}",
      journal = {Journal of Computational Physics},
         year = 1999,
        month = sep,
       volume = {154},
       number = {2},
        pages = {284-309},
          doi = {10.1006/jcph.1999.6299},
       adsurl = {https://ui.adsabs.harvard.edu/abs/1999JCoPh.154..284P}
}

@article{reville2020,
doi = {10.3847/1538-4365/ab4fef},
url = {https://doi.org/10.3847/1538-4365/ab4fef},
year = {2020},
month = {feb},
publisher = {The American Astronomical Society},
volume = {246},
number = {2},
pages = {24},
author = {Réville, Victor and Velli, Marco and Panasenco, Olga and Tenerani, Anna and Shi, Chen and Badman, Samuel T. and Bale, Stuart D. and Kasper, J. C. and Stevens, Michael L. and Korreck, Kelly E. and Bonnell, J. W. and Case, Anthony W. and de Wit, Thierry Dudok and Goetz, Keith and Harvey, Peter R. and Larson, Davin E. and Livi, Roberto and Malaspina, David M. and MacDowall, Robert J. and Pulupa, Marc and Whittlesey, Phyllis L.},
title = {The Role of Alfvén Wave Dynamics on the Large-scale Properties of the Solar Wind: Comparing an MHD Simulation with Parker Solar Probe E1 Data},
journal = {The Astrophysical Journal Supplement Series}
}

@ARTICLE{scott2018,
       author = {{Scott}, Roger B. and {Pontin}, David I. and {Yeates}, Anthony R. and
         {Wyper}, Peter F. and {Higginson}, Aleida K.},
        title = "{Magnetic Structures at the Boundary of the Closed Corona: Interpretation of S-Web Arcs}",
      journal = {Astrophys.~J.},
         year = "2018",
        month = "Dec",
       volume = {869},
       number = {1},
          eid = {60},
        pages = {60}
}

@ARTICLE{scott2021,
       author = {{Scott}, Roger B. and {Pontin}, David I. and {Antiochos}, Spiro K. and {DeVore}, C. Richard and {Wyper}, Peter F.},
        title = "{The Dynamic Formation of Pseudostreamers}",
      journal = {Astrophys.~J.},
         year = 2021,
        month = may,
       volume = {913},
       number = {1},
          eid = {64},
        pages = {64},
          doi = {10.3847/1538-4357/abec4f}
}

@ARTICLE{shoda2019,
       author = {{Shoda}, Munehito and {Suzuki}, Takeru Ken and {Asgari-Targhi}, Mahboubeh and {Yokoyama}, Takaaki},
        title = "{Three-dimensional Simulation of the Fast Solar Wind Driven by Compressible Magnetohydrodynamic Turbulence}",
      journal = {\apjl},
         year = 2019,
        month = jul,
       volume = {880},
       number = {1},
          eid = {L2},
        pages = {L2},
          doi = {10.3847/2041-8213/ab2b45},
archivePrefix = {arXiv},
       eprint = {1905.11685},
 primaryClass = {astro-ph.SR},
       adsurl = {https://ui.adsabs.harvard.edu/abs/2019ApJ...880L...2S}
}

@ARTICLE{sokolov2013,
       author = {{Sokolov}, Igor V. and {van der Holst}, Bart and {Oran}, Rona and {Downs}, Cooper and {Roussev}, Ilia I. and {Jin}, Meng and {Manchester}, IV, Ward B. and {Evans}, Rebekah M. and {Gombosi}, Tamas I.},
        title = "{Magnetohydrodynamic Waves and Coronal Heating: Unifying Empirical and MHD Turbulence Models}",
      journal = {\apj},
         year = 2013,
        month = feb,
       volume = {764},
       number = {1},
          eid = {23},
        pages = {23},
          doi = {10.1088/0004-637X/764/1/23}
}

@ARTICLE{Suess96,
       author = {{Suess}, S.~T. and {Wang}, A. -H. and {Wu}, S.~T.},
        title = "{Volumetric heating in coronal streamers}",
      journal = {\jgr},
         year = 1996,
        month = sep,
       volume = {101},
       number = {A9},
        pages = {19957-19966},
          doi = {10.1029/96JA01458},
       adsurl = {https://ui.adsabs.harvard.edu/abs/1996JGR...10119957S}
}

@ARTICLE{Singh25,
       author = {{Singh}, Talwinder and {Sankar Maity}, Samriddhi and {Chatterjee}, Piyali and {Pogorelov}, Nikolai},
        title = "{Combining local and global magnetohydrodynamic simulation frameworks to understand the evolution of coronal mass ejections}",
      journal = {Journal of Astrophysics and Astronomy},
         year = 2025,
        month = nov,
       volume = {46},
       number = {2},
          eid = {87},
        pages = {87},
          doi = {10.1007/s12036-025-10114-3},
       adsurl = {https://ui.adsabs.harvard.edu/abs/2025JApA...46...87S}
}

@ARTICLE{tokumaru2021,
       author = {{Tokumaru}, Munetoshi and {Fujiki}, Ken'ichi and {Kojima}, Masayoshi and {Iwai}, Kazumasa},
        title = "{Global Distribution of the Solar Wind Speed Reconstructed from Improved Tomographic Analysis of Interplanetary Scintillation Observations between 1985 and 2019}",
      journal = {\apj},
         year = 2021,
        month = nov,
       volume = {922},
       number = {1},
          eid = {73},
        pages = {73},
          doi = {10.3847/1538-4357/ac1862}
}

@ARTICLE{vanderholst2014,
       author = {{van der Holst}, B. and {Sokolov}, I.~V. and {Meng}, X. and {Jin}, M. and {Manchester}, IV, W.~B. and {T{\'o}th}, G. and {Gombosi}, T.~I.},
        title = "{Alfv{\'e}n Wave Solar Model (AWSoM): Coronal Heating}",
      journal = {Astrophys.~J.},
         year = 2014,
        month = feb,
       volume = {782},
       number = {2},
          eid = {81},
        pages = {81},
          doi = {10.1088/0004-637X/782/2/81}
}

@ARTICLE{vanleer1974,
       author = {{van Leer}, Bram},
        title = "{Towards the Ultimate Conservation Difference Scheme. II. Monotonicity and Conservation Combined in a Second-Order Scheme}",
      journal = {Journal of Computational Physics},
         year = 1974,
        month = mar,
       volume = {14},
       number = {4},
        pages = {361-370},
          doi = {10.1016/0021-9991(74)90019-9},
       adsurl = {https://ui.adsabs.harvard.edu/abs/1974JCoPh..14..361V}
}

@ARTICLE{wilkins2025,
       author = {{Wilkins}, Chloe P. and {Pontin}, David I. and {Yeates}, Anthony R. and {Antiochos}, Spiro K. and {Schunker}, Hannah and {Lamichhane}, Bishnu},
        title = "{The Sun's Open{\textendash}Closed Flux Boundary and the Origin of the Slow Solar Wind}",
      journal = {Astrophys.~J.},
         year = 2025,
        month = jun,
       volume = {985},
       number = {2},
          eid = {190},
        pages = {190},
          doi = {10.3847/1538-4357/adcd65}
}

@ARTICLE{wyper2022,
       author = {{Wyper}, Peter F. and {DeVore}, C.~R. and {Antiochos}, S.~K. and {Pontin}, D.~I. and {Higginson}, Aleida K. and {Scott}, Roger and {Masson}, Sophie and {Pelegrin-Frachon}, Theo},
        title = "{The Imprint of Intermittent Interchange Reconnection on the Solar Wind}",
      journal = {\apjl},
         year = 2022,
        month = dec,
       volume = {941},
       number = {2},
          eid = {L29},
        pages = {L29},
          doi = {10.3847/2041-8213/aca8ae},
       adsurl = {https://ui.adsabs.harvard.edu/abs/2022ApJ...941L..29W}
}

@ARTICLE{xia2018,
       author = {{Xia}, C. and {Teunissen}, J. and {El Mellah}, I. and {Chan{\'e}}, E. and {Keppens}, R.},
        title = "{MPI-AMRVAC 2.0 for Solar and Astrophysical Applications}",
      journal = {\apjs},
         year = 2018,
        month = feb,
       volume = {234},
       number = {2},
          eid = {30},
        pages = {30},
          doi = {10.3847/1538-4365/aaa6c8},
archivePrefix = {arXiv},
       eprint = {1710.06140},
 primaryClass = {astro-ph.SR},
       adsurl = {https://ui.adsabs.harvard.edu/abs/2018ApJS..234...30X}
}

@ARTICLE{wu2026,
       author = {{Wu}, Ziqi and {He}, Jiansen and {Hou}, Chuanpeng and {Duan}, Die and {Huang}, Jia and {Rouillard}, Alexis P. and {Verscharen}, Daniel and {Chen}, Yao and {Zhuo}, Rui and {Chen}, Tianhang},
        title = "{Multiscale Magnetic Reconnection in the Genesis of Young Slow Solar Wind}",
      journal = {\apjs},
         year = 2026,
        month = jan,
       volume = {282},
       number = {1},
          eid = {4},
        pages = {4},
          doi = {10.3847/1538-4365/ae1472},
       adsurl = {https://ui.adsabs.harvard.edu/abs/2026ApJS..282....4W}
}
\bibliographystyle{aasjournalv7}



\end{document}